\documentclass[onecolumn]{IEEEtran}
\usepackage[utf8]{inputenc}
\usepackage{graphicx}
\usepackage{amsmath}
\usepackage{amssymb}
\usepackage{mathtools}
\usepackage{multicol}
\usepackage{wrapfig}
\usepackage{color}
\usepackage{sidecap}
\usepackage{hyphenat}
\usepackage{tabularx}
\usepackage{supertabular}
\usepackage{colortbl}
\usepackage{marvosym}
\usepackage{textcomp}
\usepackage{lscape}
\usepackage{epic}
\usepackage{multicol}
\usepackage{multirow}
\usepackage{hhline}
\usepackage{pifont}
\usepackage{lastpage}
\usepackage{paralist}
\usepackage{enumitem}
\usepackage{parskip}
\usepackage{braket}
\usepackage{datetime}
\usepackage{dsfont}
\usepackage{bm}
\usepackage{hyperref}
\usepackage[hang,flushmargin]{footmisc}
\usepackage[sort,numbers,compress]{natbib}
\sidecaptionvpos{figure}{t}
\usepackage{tikz}
\usetikzlibrary{shapes.arrows,shadows}
\usetikzlibrary{decorations.pathmorphing, positioning, shapes,arrows,arrows.meta}

\newcommand{\cmark}{\ding{51}}%
\newcommand{\xmark}{\ding{55}}%

\definecolor{MyBackground}{RGB}{220,220,240}
\definecolor{MyBlue}{RGB}{0 70 128}
\definecolor{MyRed}{RGB}{166 42 42}
\definecolor{topcolor}{RGB}{202 225 255}
\definecolor{bottomcolor}{RGB}{255 255 255}
\definecolor{DeepSkyBlue}{RGB}{0,104,139}
\colorlet{MyGreen}{black!40!green}
\definecolor{MyBlue}{RGB}{0,74,153}
\definecolor{WappenBlau}{RGB}{87,129,189}
\definecolor{UniRot}{RGB}{193,0,42}
\definecolor{HellGrun}{RGB}{46,212,0}

\usepackage{tabularx,ragged2e,booktabs}
\newcolumntype{C}{>{\centering\arraybackslash}X}

\newcommand{\dd}{\mathrm{d}}

\title{Detecting a target with quantum entanglement}
\author{\IEEEauthorblockN{Giacomo Sorelli\IEEEauthorrefmark{1}\IEEEauthorrefmark{2}\IEEEauthorrefmark{3}, Nicolas Treps\IEEEauthorrefmark{1}, Fr\'ed\'eric Grosshans\IEEEauthorrefmark{2} and Fabrice Boust\IEEEauthorrefmark{3}}\\
\IEEEauthorblockA{\IEEEauthorrefmark{1}Laboratoire Kastler Brossel, Sorbonne Universit\'e, ENS-Universit\'e PSL,
Coll\`ege de France, CNRS; 4 place Jussieu, F-75252 Paris, France}
\IEEEauthorblockA{\IEEEauthorrefmark{2}Sorbonne Universit\'e, CNRS, LIP6, F-75005 Paris, France}
\IEEEauthorblockA{\IEEEauthorrefmark{3}DEMR, ONERA, Université Paris Saclay, F-91123, Palaiseau, France}}
\begin{document}
\maketitle

\begin{abstract}
In the last decade a lot of research activity focused on the use of quantum entanglement as a resource for remote target detection, i.e.\@ on the design of a quantum radar.
The literature on this subject uses tools of quantum optics and quantum information theory, and therefore it often results obscure to radar engineers.
This review has been written with the purpose of removing this obscurity. 
As such, it contains a review of the main advances in the quantum radar literature accompanied by a thorough introduction of the quantum optics background necessary for its understanding.
\end{abstract}

\tableofcontents

\section{Introduction}
The idea that by interacting with a quantum system an observer could obtain information about another system completely separated from the first one was first presented by Einstein, Podolsky and Rosen (EPR) in their seminal paper in 1935 \cite{EPR:1935}.
Later that year, Schr\"odinger called this phenomenon entanglement, and said about it that it is not ``{\it one}  but rather {\it the} the characteristic trait of quantum mechanics'' \cite{Schrodinger_en_1935,Schrodinger_de_1935}.
Nowadays, almost hundred years later, entanglement is not only considered a fundamental aspect of quantum theory, but also a precious resource for several quantum information protocols such as quantum computing, quantum communication and quantum metrology.

Recently, a whole lot of research activities focused on the use of entangled radiation in target detection, leading to the study of quantum lidars and radars.
The aim of this review is to introduce these recent advances in the theory of target detection using quantum radiation to readers with no background in quantum optics, with a particular focus on their applicability in realistic scenarios.

To achieve this goal, we will start from an historical overview to give the reader an idea about how this subject appeared and developed during the last decade in the realm of quantum information.
Then, before jumping into a quite detailed description of the quantum target-detection protocol, also known as quantum illumination, we will introduce some preliminary concepts.
We will first present some basics of quantum optics, then we will discuss how to decide between two hypotheses (target present/target absent) by distinguishing different probability distributions (a subject that should be familiar to radar engineers) and adapt these concepts to the problem of discriminating different quantum states.

After having set this background, we will discuss the performances of quantum illumination and point out under which conditions one can obtain a quantum advantage.
We will also point out the main practical limitations of these protocols, most of which were already reported in \cite{pirandola2018, ShapiroReview2019}.  
In particular, we will see that a quantum advantage is possible only when extremely low signal powers are employed, and therefore quantum illumination is of very little use for remote target detection. 

We will conclude this review by giving a quick overview of the state of the art of quantum radar experiments. 
In particular, we will discuss how, in most cases, these experiments are unable to prove the quantum advantage predicted by quantum illumination theory.

\section{A historical overview}
The history of quantum illumination started in 2008, following two lines of research.
The works \cite{lanzagorta2010quantum,lanzagorta2011quantum} considered the radar problem from a quantum interferometry perspective. 
However, these works considered highly idealized scenarios, and neglected the influence of thermal background. 
Since this review is focused on the practicality of quantum radars, we will not further discuss this approach, and focus on the other approach pioneered by Seth Lloyd the same year \cite{Lloyd_Science_2008}, when he studied how to use quantum light to detect a weakly reflecting target embedded in thermal background \cite{Lloyd_Science_2008} \footnote{A reader interested in the interferometric approach to quantum radar is referred to the original papers, or to the review \cite{torrome2020introduction}}.
In his work, Lloyd considered two protocols: the first interrogates the target region with $N$ independent single photons, while the second protocol uses $N$ photons are all entangled with one another. 
Lloyd's results showed that in the entanglement-based protocol the probability of making a wrong decision about the presence of the target is dramatically lower than in the single-photons' one.
These results were welcomed with excitation from the quantum optics community because they seemed to suggest that entanglement could revolutionize current radar technology.

However, the reader should be aware that both protocols proposed by Lloyd are actually quantum, accordingly the work in \cite{Lloyd_Science_2008} does not prove that a quantum radar can outperform any classical radar.
On the contrary, both Lloyd's protocols perform worse than a detection scheme using coherent states \cite{Shapiro_NJP_2009}, that as we will explain in Sec.\@ \ref{sec:gaussian} can be considered the quantum mechanical description of a traditional radar.
Fortunately, Tan {\it et al.}\@ \cite{Tan_PRL_2008} had already proposed a more sophisticated quantum-illumination protocol that, in a very specific regime, can provide a quantum advantage.
This review aims at presenting this protocol, its consequences and limitations.

In their works on quantum illumination Lloyd \cite{Lloyd_Science_2008} and Tan {\it et al.\@} \cite{Tan_PRL_2008} considered the problem of discriminating between two hypotheses: target present and target absent (see section \ref{sec:discrimination}). 
In this setting, by using tools from quantum information theory (see section \ref{sec:Bayes}), these authors determined the lowest probability of choosing the wrong hypothesis. 
In particular, Tan {\it et al.\@} \cite{Tan_PRL_2008} proved that the lowest error probability achievable with entangled radiation is given by 
\begin{equation}
P_e^{QI} \approx \tfrac{1}{2}e^{-M\kappa N_s/N_B},
\end{equation}
where $M$ is the number copies of the quantum state used to probe the target, $N_s$ is the average number of signal photons in a single copy of the quantum state, and $N_B$ is the average number of background photons present when the return from a single-copy transmission is detected, $\kappa$ is the roundtrip radar-to-target-to-radar transmissivity, such that $\kappa N_s$ is the number of photons coming back to the receiver when the target is present.
As we will abundantly discuss in this review, this expression requires the usual assumptions of strong thermal background $N_B \gg 1$ and low roundtrip transmissivity $\kappa \ll 1$, but also a low signal assumptions $N_S \ll 1$. 
Accordingly, in order to have a satisfactory signal-to-noise ratio $SNR = M \kappa N_S/ N_B$, $M \gg 1$ is required. 
In the best case of classical illumination, one has instead
\begin{equation}
P_e^{CS} \approx \tfrac{1}{2}e^{-M\kappa N_s/4N_B}
\end{equation}
Therefore, following the protocol of Tan {\it et al.\@} \cite{Tan_PRL_2008}, the minimal error probability exponent achievable in the quantum case is $6$ dB smaller than the best obtainable in the classical case.
This quantum advantage stems from entanglement which provides stronger-than-classical correlations between the signal transmitted to the target region and an idler pulse retained in the radar detector. 
However, these peculiar correlations cannot be detected with standard receivers. 
Accordingly, at the time of the work \cite{Tan_PRL_2008}, December 2008, no explicit detection scheme able to achieve this lower bound was known.

The first receivers able to obtain a quantum advantage in a quantum illumination scheme were proposed in November 2009 by Guha and Erkmen \cite{Guha_PRA_2009}.
On the one hand, these two receivers have the advantage of being implementable with current technology; on the other hand they do not achieve the ultimate limit allowed by quantum mechanics as discussed by Tan \emph{et al.\@}, but they only provide a $3$ dB advantage.
A receiver able to achieve this ultimate bound has been found in 2017 only by Zhuang, Zhang and Shapiro \cite{Zhuang_PRL_2017}.
Such a receiver is extremely complicated and far beyond the capability of state of the art experiments.
However, knowing its blueprint has great theoretical relevance since it allows to determine the receiving operating characteristic (ROC), namely the probability of target detection as a function of the false alarm probability (see section \ref{sec:Neyman--Pearson}).

The attention of the radar community was caught in 2015 when the first theoretical proposal for performing quantum illumination with microwaves \cite{BarzanjehPRL2015} appeared.
This work suggested to use an electro-optomechanical device to produce a microwave signal beam entangled with an optical idler beam that is stored for subsequent joint measurement with the radiation returned from the region under interrogation.
This first idea had the merit to open quantum illumination research to microwaves, which are more suitable for target detection, but it has never been used in experimental implementations.
However, the same work also suggested to use more practical superconducting devices, such as the Josephson parametric amplifiers (JPA). Such devices have been indeed used in several experiments thereafter \cite{luong2019quantum,luong2019receiver,Chang_applied_phys_lett_2019, Barzanjeh_arxiv_2019}.

Regarding the experimental implementations, it is worth mentioning that the only work demonstrating the quantum advantage of the Tan {\it et al.\@} protocol \cite{Tan_PRL_2008} with the Guha-Erkmen receiver was performed at optical frequencies in 2015 \cite{Zhang_PRL_2015}.
Several experiments in the microwave regime have been reported in the last few years \cite{luong2019quantum,luong2019receiver,Chang_applied_phys_lett_2019, Barzanjeh_arxiv_2019}. 
However, in the attempt to work around some technical difficulties, all these experiments modified the quantum illumination protocol and ended up in regimes where it can be proved that quantum entanglement does not provide any quantum advantage \cite{ShapiroReview2019}.
 
\section{Preliminaries}
In this section, we provide the background knowledge needed to understand quantum illumination.
More specifically, we deal with two important subjects: the quantum mechanical description of the electromagnetic field and the problem of discriminating between two different quantum states.

\subsection{Quantum optics crash course}
This section is particularly intended for radar engineers with familiarity with classical electromagnetic theory, but no background in quantum optics.
We will start by considering the standard procedure to quantize the electromagnetic field \cite{loudon,agarwal_2012}.
We will then introduce those quantum states that are used to describe classical radiation and thermal noise in quantum optics. 
Finally, the concept of entanglement, and the particular entangled state used in quantum illumination will be presented.
Readers familiar with quantum optics can safely skip to Sec.\@\@ \ref{sec:discrimination}.

\subsubsection{Quantization of the electromagnetic field}
Let us start by defining a set of {\it modes of the electromagnetic field} \cite{Fabre_2019} $\left\lbrace{\bf f }_i({\bf r},t)\right\rbrace$ as solutions of the wave equation 
\begin{equation}
\left( \nabla^2 -\frac{1}{c^2}\frac{\partial^2}{\partial t^2}\right){\bf f }_i({\bf r},t) = 0,
\label{wave_eq}
\end{equation}
satisfying the transversality and orthonormality conditions
\begin{subequations}
\begin{align}
&\nabla \cdot {\bf f }_i({\bf r},t) = 0,\\
&\frac{1}{V}\int{\bf f }^*_i({\bf r},t) {\bf f }_j({\bf r},t) \dd^3 {\bf r} =\delta_{i,j},
\end{align}
\end{subequations}
where $^*$ denotes complex conjugation, $\delta_{i,j}$ is the Kronecker delta function, and $V$ is an arbitrarily large volume containing the full physical system under consideration.
Given an arbitrary orthonormal mode basis $\left\lbrace{\bf f }_i({\bf r},t)\right\rbrace$, any solution of Maxwell's equations, e.g the electric field ${\bf E}({\bf r},t)$, can be expressed (in SI unit) as
\begin{equation}
{\bf E}({\bf r},t) = \sum_{i=1}^N \left(\frac{\hbar \omega_i}{2\epsilon_0 V } \right)^{1/2} \left(a _i{\bf f }_i({\bf r},t)  + a _i^*{\bf f }^ *_i({\bf r},t)\right),
\label{E_expansion}
\end{equation}
with $\omega_i$ the central frequency associated with the quasi-monochromatic modes $\bf{f}_i$ and $a_i = (q_i + i p_i)/2$ are complex numbers, where $q_i$ and $p_i$, are known as {\it fields quadratures}.\footnote{In this review, we will use the typical notation of the quantum optics literature $a_i = (q_i + i p_i)/2$. 
However, in the radar literature it is more common to use the notation $a_i = (I_i + i Q_i)/2$.
We invite those reader more familiar with the second notation to be aware of the change of meaning of the letter "q".
}

We can now quantize the field ${\bf E}({\bf r},t)$ by replacing the coefficients in Eq.\@ \eqref{E_expansion} with operators according to
\begin{subequations}
\begin{align}
a_i &\to \hat{a}_i,\\
a^*_i &\to \hat{a}^\dagger_i,
\end{align}
\end{subequations}
where the $^\dagger$ denotes Hermitian conjugation. The operators $\hat{a}^\dagger_i$ and $\hat{a}_i$ are known  as {\it creation} and {\it annihilation} operators and take their names from their action on {\it photon-number states}.
In fact, if we denote with $\ket{n_i}_i$ the quantum state of the electromagnetic field containing $n_i$ photons whose spatial and temporal profile is defined by the mode ${\bf f }_i({\bf r},t)$, the action of the annihilation (creation) operator on such a state is to remove (add) a photon:
\begin{subequations}
\begin{align}
\hat{a}_i\ket{n_i}_i &= \sqrt{n}\ket{n_i-1}_i, \label{a}\\
\hat{a}^\dagger_i \ket{n_i}_i. &= \sqrt{n+1}\ket{n_i+1}_i\label{a_dag}.
\end{align}
\label{photon_numb_a}
\end{subequations}
It follows directly from Eqs.\@ \eqref{photon_numb_a} that $\hat{a}^\dagger_i \hat{a}_i\ket{n_i}_i = n_i \ket{n_i}_i$. As a consequence, we define $\hat{n_i} = \hat{a}^\dagger_i \hat{a}_i$, and we call it the {\it photon number operator}.
Moreover, creation and annihilation operators obeys the following commutation relations
\begin{equation}
\left[ \hat{a}_i, \hat{a}^\dagger_j\right] = \hat{a}_i \hat{a}^\dagger_j - \hat{a}_j^\dagger \hat{a}_i  =\delta_{i,j}.
\label{comm}
\end{equation} 

Given the above definitions, we can write the quantum mechanical operator describing the electromagnetic field as 
\begin{equation}
{\bf \hat{E}}({\bf r},t) = \sum_{i=1}^N \left(\frac{\hbar \omega_i}{2\epsilon_0 V } \right)^{1/2} \left(\hat{a} _i{\bf f }_i({\bf r},t)  +\hat{a} _i^\dagger{\bf f }^ *_i({\bf r},t)\right).
\label{Quantum_E_expansion}
\end{equation}
Eq.\@ \eqref{Quantum_E_expansion} is the conceptual core of {\it quantum optics}, where the {\it quantum} part comes from the creation and annihilation operators, their action on photon-number states \eqref{photon_numb_a}, and their commutation relation \eqref{comm}. 
On the other hand, the {\it optics} part stems from the mode functions ${\bf f }_i({\bf r},t)$, that are solutions of the wave equation \eqref{wave_eq}, and therefore they evolve and propagate as classical electromagnetic waves.

After having defined the quantum-mechanical electromagnetic field operator, we now introduce the formalism to describe quantum states of light. 
Let us start with the {\it vacuum state} $\ket{0}$, namely the state of the electromagnetic field containing no photons. 
When no photons are present, no photon can be removed ($\hat{a}_i\ket{0} = 0$). 
On the contrary, photons can be added to the vacuum, and we can obtain the states $\ket{n_i}_i$ by applying $n_i$ times Eq.\@ \eqref{a_dag}: $\ket{n_i} = \left(\hat{a}_i^\dagger\right)^{n_i}\ket{0}/(n_i!)$.
The most general {\it pure} state of the total field is then given by a linear superposition of products of photon-number states of the individual modes \cite{loudon}, 
\begin{equation}
\ket{\Psi} = \sum_{n_1} \cdots \sum_{n_m} \cdots C_{n_1\cdots n_m \cdots } \ket{n_1}_1 \otimes \dots \otimes\ket{n_m}_m\otimes\cdots,
\label{pure}
\end{equation}
where the symbol $\otimes$ denotes the tensor product.

In the linear superposition in Eq.\@ \eqref{pure}, the phase relation between the different multi-mode photon-number states in $\ket{\Psi}$ are perfectly defined by the complex coefficients $C_{n_1\cdots n_m \cdots }$.
However, knowing these phase relations is not always possible in quantum optics, and sometimes the only thing we can specify are a set of probabilities for the field to be found in certain states. 
When this is the case, the state of the field cannot be written as a {\it pure state}, i.e.\@ it is not given by Eq.\@ \eqref{pure}, and we refer to it as a {\it statistical mixture} or simply as a {\it mixed state}.
A mixed state can be expressed as a {\it density operator} \cite{loudon}
\begin{equation}
\rho = \sum_i P_i \ket{\psi_i}\bra{\psi_i},
\label{density-matrix}
\end{equation}
where $P_i$ represents the probability for the field to be found in the pure state $\ket{\psi_i}$.
Accordingly, we have $0 \leq P_i\leq 1$ and $\sum_i P_i = 1$. 
The mean value of a quantum mechanical operator $\hat{O}$ can then be expressed as 
\begin{equation}
\langle \hat{O} \rangle = \rm{tr} \left(\hat{O} \rho\right) = \sum_i P_i \bra{\psi_i} \hat{O}\ket{\psi_i},
\label{mean}
\end{equation}
where $\rm{tr}$ denotes the trace operation.
Eq.\@ \eqref{mean} can be understood as the sum of the expectation values $\bra{\psi_i} \hat{O}\ket{\psi_i}$ of the operator $\hat{O}$ for the states $\ket{\psi_i}$ weighted with the probabilities of these states to appear in the statistical mixture $\rho$.

Let us conclude by pointing out that a pure state is a special mixed state for which a particular probability $P_i =1$, while all others vanish, i.e.\@ $\rho = \ket{\psi}\bra{\psi}$.
Therefore, when dealing with general expressions valid both for mixed and pure states, especially when dealing with mean values (see Eq.\@ \eqref{mean}), it is often convenient to use the density matrix formalism.

\subsubsection{Phase space distribution and Gaussian states}
\label{sec:gaussian}
In this section, we will discuss the properties of the particular quantum states of light which are relevant for quantum illuminations. 
All these states fall in the class of the so called {\it Gaussian states}, namely quantum states which are fully characterized by the first and second moments of the quadrature operators \cite{Weedbrook_Rev_Mod_Phys_2012}.
In particular, mixed Gaussian states are the quantum analogue of classical Gaussian noise.
Those states are better understood in terms of their {\it Wigner function}, which we will introduce in the following.

{\bf Wigner function}

Let us consider a physical system consisting of $N$ electromagnetic modes, we can arrange the corresponding quadrature operators $\hat{q}_i = \hat{a}_i + \hat{a}_i^\dagger$ and $\hat{p}_i = i(\hat{a}^\dagger_i - \hat{a}_i$) in the operators vector 
\begin{equation}
{\bf \hat{x}} = \left( \hat{q}_1, \hat{p}_1, \cdots, \hat{q}_N, \hat{p}_N\right), 
\end{equation}
and define the operator 
\begin{equation}
D({\bm \xi}) = \exp \left( i {\bf \hat{x}}^T {\bm \Omega} {\bm \xi} \right),
\end{equation}
where $^T$ denotes the transpose,  ${\bm \xi} \in \mathbb{R}^{2N}$, and 
\begin{equation}
{\bm \Omega} = \bigoplus_{i=1}^N{\bm \omega} = 
\begin{pmatrix}
{\bm \omega}& & \\
&\ddots &\\
& &{\bm \omega}
\end{pmatrix},
\quad \rm{with} \; 
{\bm \omega} = 
\begin{pmatrix}
0& 1&\\
-1& 0
\end{pmatrix},
\end{equation}
is $2N \times 2N$ matrix known as the symplectic form. 
The density operator $\rho$ of an arbitrary quantum state is then equivalent to the characteristic function \cite{Weedbrook_Rev_Mod_Phys_2012}
\begin{equation}
\chi({\bm \xi}) = {\rm tr} \left[ \rho D({\bm \xi}) \right], 
\end{equation}
and to its Fourier transform
\begin{equation}
W({\bm x }) =  \int_{\mathbb{R}^{2N}} \frac{\dd^{2N} {\bm \xi}}{(2\pi)^{2N}} \exp \left( - i {\bf x}^T {\bm \Omega}  {\bm \xi} \right)\chi({\bm \xi}),
\end{equation}
which is a normalized, but generally non-positive, {\it quasi-probability} distribution known as the {\it Wigner function} \cite{Weedbrook_Rev_Mod_Phys_2012}.
The continuous variables ${\bm x }$ are the eigenvalues of the quadrature operators, and they span a real $2N-$dimensional space known as {\it phase space}.\footnote{To be more precise the phase space is the symplectic space formed by $\mathbb{R}^{2N}$ equipped with the symplectic form ${\bm \Omega}$ \cite{Weedbrook_Rev_Mod_Phys_2012}. Such a mathematical subtlety has several relevant consequences in quantum optics, and some of them play quite an important role also in quantum illumination. However, for the sake of simplicity, in this review we will avoid to refer explicitly to these rather involved mathematical details, but we will refer the interested readers to the relevant literature \cite{dutta1995real}.}
Even though, the Wigner function itself is a quasi-probability distribution, its {\it marginals} corresponding to measurable quadratures --- like ${\bf q}$, ${\bf p}$ and some linear combinations of their coordinates --- are proper measurement probabilities.
For example, the probability distribution of the ${\bf q} = (x_1,x_3,\cdots,x_{2N-1})$ quadratures can be obtained from the Wigner function by integrating over all the ${\bf p} =(x_2,x_4, \cdots, x_{2N})$ quadratures 
\begin{equation}
P({\bf q}) = \int W({\bf x}) \left(\prod_{n=1}^N\dd x_{2n}\right).
\end{equation}

A Wigner function can be characterized by the statistical moments of the corresponding quantum state. 
In particular, the first two moments are the {\it mean vector}
\begin{equation}
\bar{\bf x} = \langle {\bf x} \rangle = \rm{tr} \left(\rho {\bf x} \right),
\end{equation}
and the {\it covariance matrix} ${\bf V}$, whose elements are defined as
\begin{equation}
V_{i,j} = \frac{1}{2}\langle\lbrace\Delta \hat{x}_i,\Delta \hat{x}_j \rbrace \rangle,
\end{equation}
where the curly brackets denote the anti-commutator ($\lbrace \hat{A}, \hat{B}\rbrace = \hat{A}\hat{B} + \hat{B}\hat{A}$), and $\Delta \hat{x}_i = \hat{x}_i - \langle \hat{x}_i \rangle$. 
The diagonal elements of this matrix represent the covariances of the quadrature operators $V_{ii} = V(\hat{x}_i) = \langle (\Delta \hat{x}_i)^2 \rangle = \langle \hat{x}_i^2 \rangle - \langle \hat{x}_i\rangle^2$, while the off-diagonal elements quantify the correlations between the different modes.

The fact that creation and annihilation operators do not commute [see Eq.\@ \eqref{comm}] implies the following {\it uncertainty relation} \cite{Weedbrook_Rev_Mod_Phys_2012}
\begin{equation}
{\bm V} + i{\bm \Omega} \geq 0,
\label{uncertainty}
\end{equation}
which must be interpreted in the matrix sense, meaning that all eigenvalues of the matrix ${\bm V} + i{\bm \Omega}$ are larger than zero.
In particular, Eq.\@ \eqref{uncertainty} tells us that 
\begin{equation}
V(\hat{q}_i)V(\hat{p}_i) \geq 1,
\label{uncertainty_qp}
\end{equation}
which is the quantum optical analogue of the famous Heisenberg uncertainty principle between position and momentum in standard quantum mechanics \cite{sakurai2011modern}, and tells us that we cannot measure both quadratures $q_i$ and $p_i$ at  the same time with arbitrary precision.

A very important family of quantum states, which comprise all states that we will consider in this manuscript, is the one of {\it Gaussian states} \cite{Weedbrook_Rev_Mod_Phys_2012}. 
These states are fully determined by their mean vector and their covariance matrix, and their Wigner function is given by the Gaussian function
\begin{equation}
W({\bm x })  = \frac{\exp\left[ - ({\bm x } -\bar{\bm x})^T{\bf V}^{-1} ({\bm x } -\bar{\bm x})/2 \right]}{(2\pi)^{N} \sqrt{\det {\bm V}}}, 
\label{Gaussian_W}
\end{equation}
where det denotes the determinant. 

The formalism described above is fairly general and allows us to describe arbitrary multi-mode quantum states of light. 
In the following, we will use it to introduce some single-mode and two-mode quantum states which are relevant for quantum illumination.

{\bf Coherent states a.k.a.\@ quasi classical states}

Let us now consider the so called {\it coherent states} $\ket{\alpha}$. 
These are single-mode states defined as the eigenstates of the creation operator \cite{loudon}
\begin{equation}
\hat{a} \ket{\alpha} =\alpha \ket{\alpha},
\label{coherent_def}
\end{equation}
with $\alpha = |\alpha|\exp (i\varphi)$ a complex number.
Equation \eqref{coherent_def} is solved by the state \cite{loudon}
\begin{equation}
\ket{\alpha} = \exp \left( -|\alpha|^2/2\right)\sum_{n=0}^{\infty} \frac{\alpha^n}{\sqrt{n!}}\ket{n},
\label{coherent}
\end{equation}
and its Wigner function takes the Gaussian form in Eq.\@ \eqref{Gaussian_W} \cite{Weedbrook_Rev_Mod_Phys_2012} with mean vector
\begin{equation}
\bar{\bf x}  =(q,p) = \left(\alpha + \alpha^*, \alpha - \alpha^*\right)/2 =  \left(|\alpha|\cos\varphi, |\alpha|\sin\varphi\right)
\end{equation} 
 and covariance matrix ${\bm V} = \mathds{1}_2$, with $\mathds{1}_2$ the $2\times 2$ identity matrix. 

A coherent state is thus a superposition of infinitely many photon-number states with a mean photon number determined by the parameter $\alpha$ according to
\begin{equation}
\langle \hat{n} \rangle = \bra{\alpha}\hat{a}^\dagger\hat{a}\ket{\alpha} = |\alpha|^2.
\label{num_coh}
\end{equation}

\begin{wrapfigure}[14]{r}{0.42\textwidth}
\vspace{-0.8 cm}
\begin{center}
\includegraphics[width=0.37\textwidth]{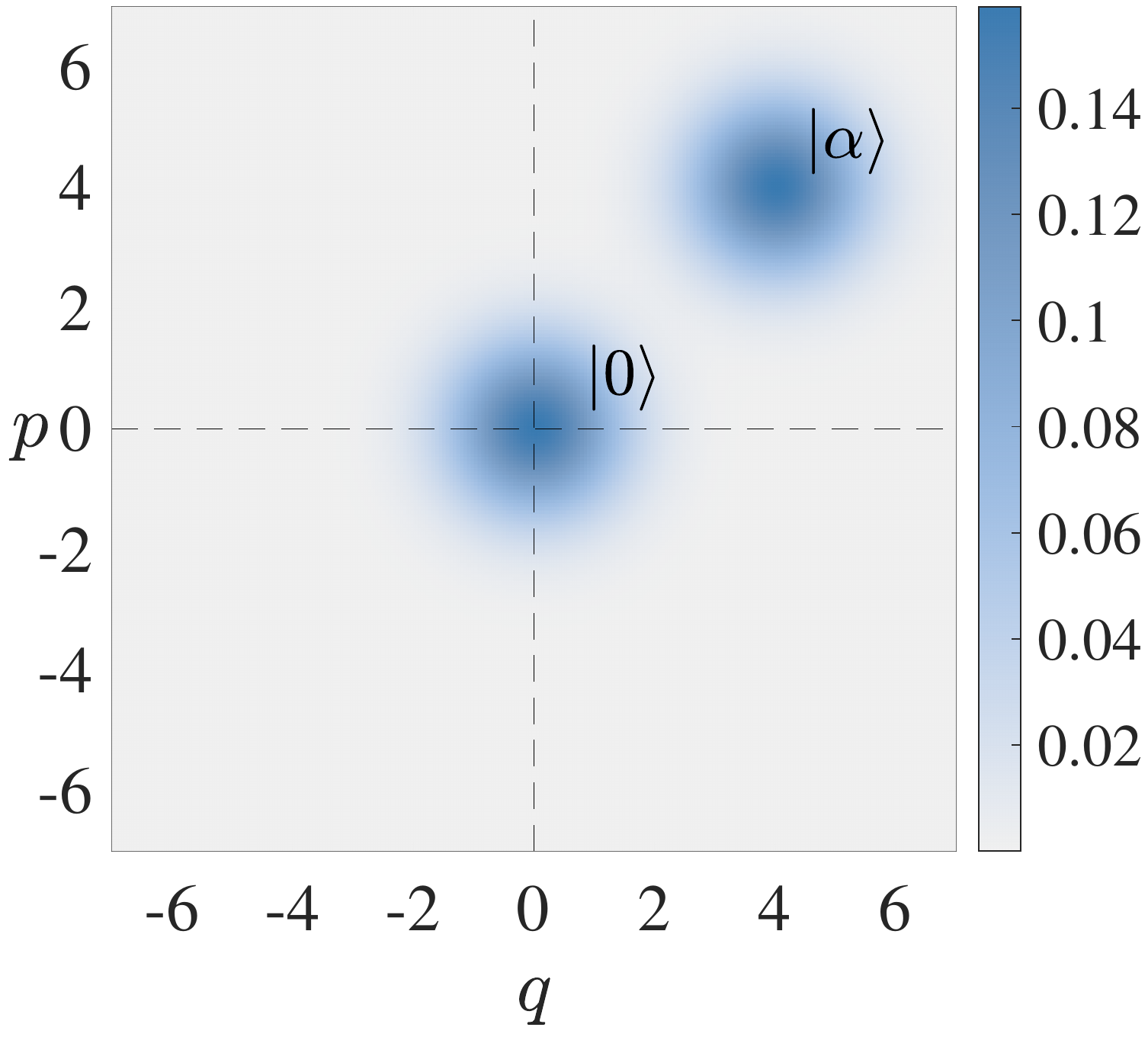}
\caption{Wigner distributions of the vacuum state $\ket{0}$ and a coherent state $\ket{\alpha}$ with $\alpha = 4+ 4i$.}
\label{Fig:coherent}
\end{center}
\end{wrapfigure}

In particular, by setting $\alpha =0$ in Eq.\@ \eqref{coherent}, we obtain the vacuum state $\ket{0}$. 
Therefore, every coherent states has the same covariance matrix as the vacuum.
Moreover, this particular form of the covariance matrix ${\bm V}$ saturates the uncertainty relation \eqref{uncertainty_qp} \cite{Weedbrook_Rev_Mod_Phys_2012}.
As a consequence, coherent states allow for the best precision in simultaneous measurements of the $q$ and $p$ quadratures.
The Wigner distribution of the vacuum and of a coherent state with non-zero photon number are compared in Fig.\@ \ref{Fig:coherent}.

Coherent states are the quantum states that most closely resemble a classical electromagnetic field. 
In fact, the mean value of the electromagnetic field operator \eqref{Quantum_E_expansion} for a coherent state is given by

\begin{equation}
\left\langle{\bm E}({\bm r},t)\right\rangle = \bra{\alpha}\hat{\bm E}({\bm r},t)\ket{\alpha} = \sqrt{\hbar \omega/2\epsilon_0 V}\left(\alpha f({\bm r},t) +  \alpha^* f^*({\bm r},t)\right),
\label{mean_coher}
\end{equation}
where we have used Eq.\@ \eqref{coherent_def}, and the fact that the state is single mode.

If we consider the particular case of a plane wave $f({\bm r},t) = \exp[-i ({\bm k}\cdot {\bm r} -\omega t)]$, we obtain
\begin{equation}
\left\langle{\bm E}({\bm r},t)\right\rangle = \sqrt{\hbar \omega/2\epsilon_0 V}|\alpha| \cos ({\bm k}\cdot {\bm r} -\omega t -\varphi).
\label{coh_plane}
\end{equation}
Let us recall that the mean energy of a quantum state of light is given by $\hbar \omega (\langle\hat{n} \rangle + 1/2)$, where $\hbar \omega /2$ is the zero-point energy of the vacuum, which manifest itself as noise when measuring the field quadratures.
By using Eq.\@ \eqref{num_coh}, we see that this term becomes negligible when $|\alpha| \gg 1$. 
In this limit, a field measurement on a coherent state results in a noiseless signal equal to $\left\langle{\bm E}({\bm r},t)\right\rangle$, which according to Eq.\@ \eqref{coh_plane} is exactly what we would expect for a classical electromagnetic signal: a cosine wave with amplitude proportional to square root of the energy carried by the wave.

To convince oneself that this is a peculiarity of coherent states, the reader should notice that for photon number states $\ket{n}$, $\bra{n}\hat{\bm E}({\bm r},t)\ket{n} =0$, for every value of $n$.
Therefore, in our discussion of quantum illumination, we will refer to coherent states every time we want to compare with classical light.

{\bf Thermal states a.k.a. the noise}

Let us now discuss another important class of single-mode Gaussian states, namely the quantum state of the black body radiation which is often associated with thermal noise.
Such a state is a statistical mixture of photon number states \cite{agarwal_2012}
\begin{equation}
\rho_T = \sum_{n=1}^{\infty} \frac{(N_T)^n}{(N_T+1)^{n+1}}\ket{n}\bra{n},
\label{thermal}
\end{equation} 
with $N_T = \rm{tr} (\hat{n} \rho)$ the mean photon number.
The mean photon number for a mode with frequency $\omega$, at a given temperature $T$ is determined by the Bose-Einstein distribution \cite{agarwal_2012}
\begin{equation}
N_T = \frac{1}{\exp[\hbar \omega/(k_B T)] -1},
\label{NT}
\end{equation}
where $k_B$ is the Boltzmann constant. 
Thermal states \eqref{thermal} describes the noise background in quantum illumination. 
In this context, $T$ is assumed to be the sky temperature at the radar receiver.

\begin{wrapfigure}[15]{r}{0.5\textwidth}
\vspace{-0.6 cm}
\begin{center}
\includegraphics[width=0.37\textwidth]{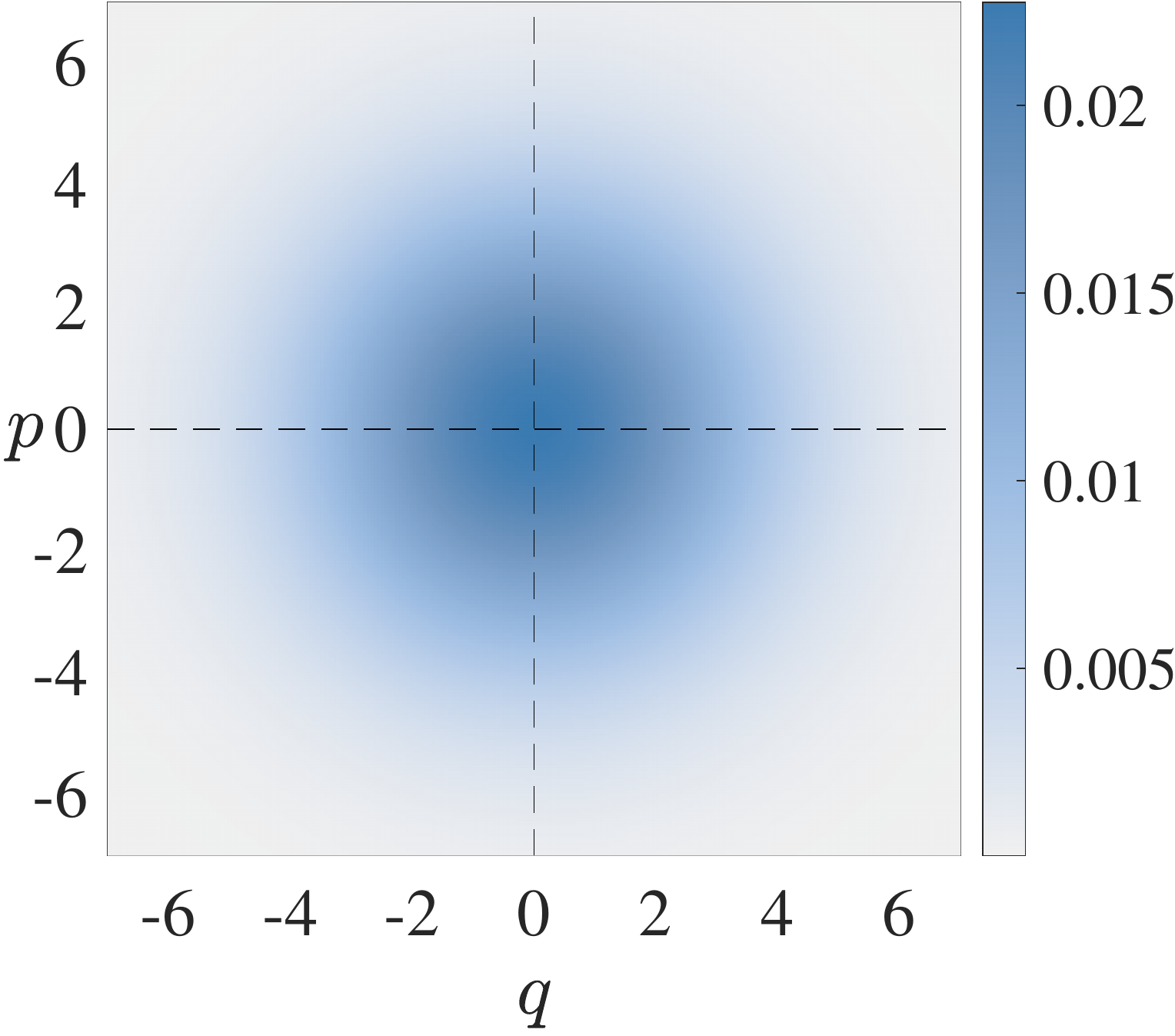}
\caption{Wigner distributions of a thermal state $\rho_T$ with mean photon number $N_T=3$.}
\label{Fig:thermal}
\end{center}
\end{wrapfigure}

Thermal states \eqref{thermal} are Gaussian states with a Wigner function of form \eqref{Gaussian_W} \cite{Weedbrook_Rev_Mod_Phys_2012} with mean vector $\bar{\bf x}  =(0,0)$ and covariance matrix ${\bm V} = (2N_T +1)\mathds{1}_2$. 
Therefore, as illustrated in Fig.\@ \ref{Fig:thermal} (note that the color scale is different with respect to the one in Fig.\@ \ref{Fig:coherent}), their phase space representation is always centred at the origin and have a width which is determined by the mean photon number $N_T$.
Accordingly, the Gaussian quasi-probability distribution associated with a thermal state with $N_T>0$ is broader than the one of a coherent state. 

Simply looking at Eq.\@ \eqref{NT} one can notice a striking difference between the intensity of noise backgrounds at optical and microwave frequencies, that will be relevant in our discussion about quantum radar. 
For optical frequencies $N_T$ is negligible, on the contrary for microwave frequencies  $N_T$ is significantly larger than one.\footnote{An estimate of the noise background based only on Eq.\@ \eqref{NT} is actually very rough, especially at optical frequency. However, a more accurate calculation taking in account parameters other than temperature, as for example the Sun irradiance, would lead to $N_T \approx 10^{-6}$ at THz frequencies and $N_T \approx 10^3$ at $100$ MHz, which does not change the substance of the above statement.}

{\bf  Two-mode squeezed  vacuum a.k.a. twin beams}

Let us now consider a quantum state involving two modes of the electromagnetic field defined by the annihilation operators $\hat{a_1}$ and $\hat{a_2}$.
A pure quantum state of a two mode field is called {\it separable} if it can be written as a product of a state of mode $a_1$ and a state of mode $a_2$: $\ket{\Psi} = \ket{\psi}_1\ket{\phi}_2$ \cite{agarwal_2012}. 
A quantum state that cannot be written in this form is called {\it entangled} \cite{agarwal_2012}.
A two-mode entangled state that is essential for quantum illumination is the {\it two-mode squeezed vacuum}:
\begin{equation}
\ket{\xi} = \sum_{n=1}^{\infty} \sqrt{\frac{(N_s)^n}{(N_s+1)^{n+1}}}\ket{n}_1\ket{n}_2,
\label{TMSV}
\end{equation}
where $N_s$ is the mean photon number in the two modes, i.e $\langle \hat{n}_1 \rangle = \langle \hat{n}_2 \rangle = N_s$. 
The state \eqref{TMSV} is also a Gaussian state \cite{Weedbrook_Rev_Mod_Phys_2012} with mean vector  $\bar{\bf x}  =(0,0,0,0)$ and covariance matrix
\begin{equation}
{\bm V} =
\begin{pmatrix}
S & 0 & C_q & 0 \\
0 & S & 0 & -C_q\\
C_q & 0 & S & 0\\
0 & -C_q & 0 & S
\end{pmatrix},
\label{cov_TMSV}
\end{equation}
where $S = 2N_s +1$ represents the variances of the quadratures of the two modes, while $C_q = 2\sqrt{N_s(N_s+1)}$ indicates the correlations between the two modes.

\begin{SCfigure}[][h]
\centering
\includegraphics[width=0.6\textwidth]{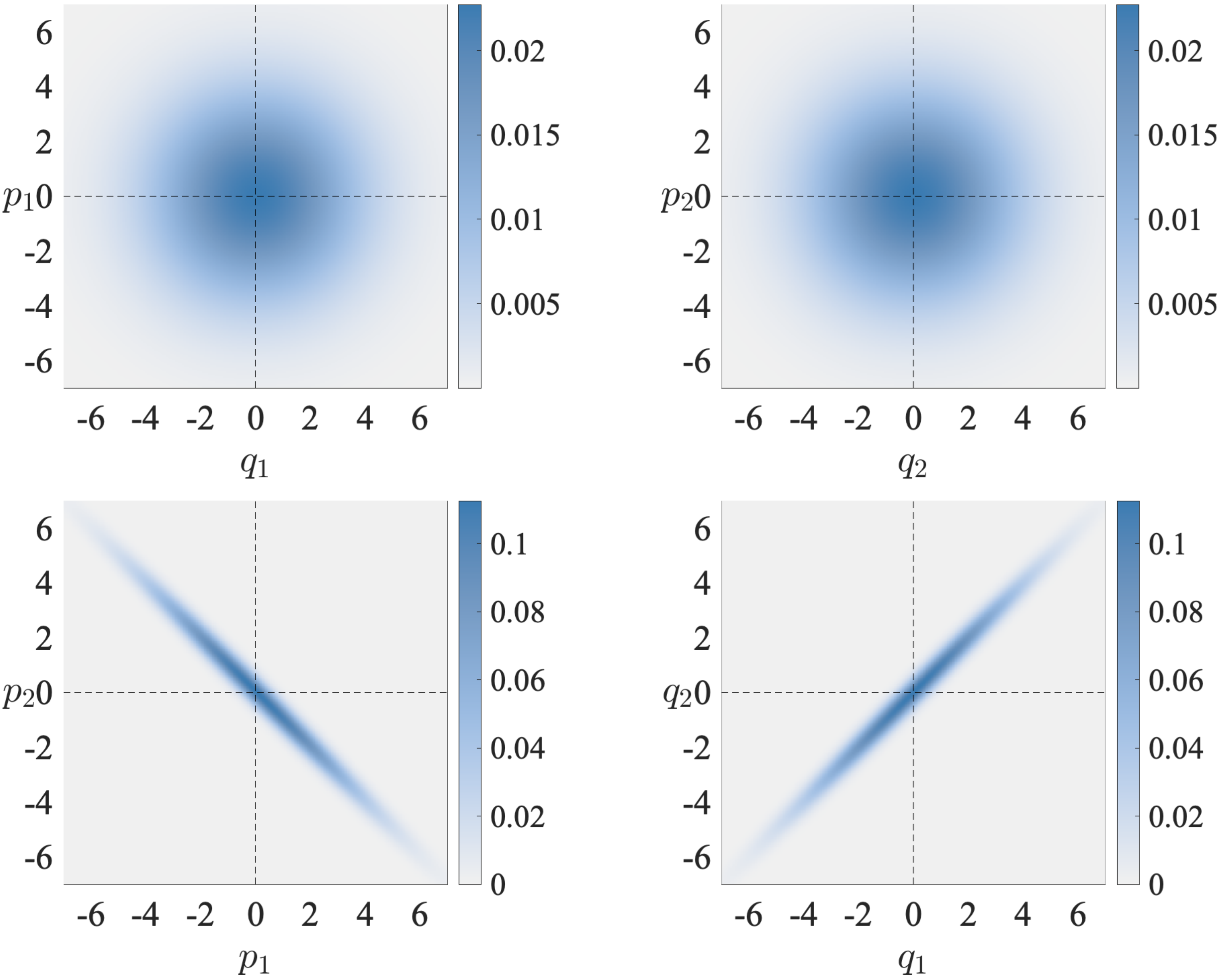}
\caption{Marginals of the Wigner function of a two-mode squeezed vacuum state with mean photon number $N_s=3$. 
The top row shows the distribution of the $q$ and $p$ quadratures of each of the two modes. 
The distributions in the bottom row demonstrates the correlation between the two modes.}
\label{Fig:TMSV}
\end{SCfigure}

In Fig.\@ \ref{Fig:TMSV}, we plotted the marginals of the two-mode squeezed vacuum Wigner functions (note that different panels have different color scales).
In the two top panels, we see that the probability distributions for the quadratures of a single mode are identical to those of a thermal state (compare with Fig.\@ \ref{Fig:thermal}), and therefore reveal no correlation.
On the other hand, in the bottom row, we see that the probability distributions $P(p_1,p_2)$ and $P(q_1,q_2)$ are strongly {\it squeezed} along a specific direction. 
In particular, the $q$ quadratures are correlated, while the $p$ quadratures are anti-correlated, we are therefore in presence of strong {\it phase-sensitive} cross-correlations.
The name two-mode squeezed vacuum stems from these cross-correlations, and in the following we will see that the strength of these correlations surpass what is allowed by classical physics.

To this goal, let us now make a few comments on the covariance matrices of two-mode Gaussian states.
In particular, in quantum illumination, we will always deal with  matrices of the form 
\begin{equation}
{\bm V} =
\begin{pmatrix}
A & 0 & C & 0 \\
0 & A & 0 & -C\\
C & 0 & B & 0\\
0 & -C& 0 & B
\end{pmatrix},
\label{cov_TM}
\end{equation}
of which \eqref{cov_TMSV}  is a particular case. 
We have seen above that the correlations between the modes are encoded in the off-diagonal elements of the covariance matrix.
Therefore, it is reasonable to expect that it exists a criterion that tells us if a Gaussian state is entangled or not by comparing the off-diagonal terms of the covariance matrix with the diagonal ones.
For a covariance matrix in the form \eqref{cov_TM}, such an entanglement criterion reads \cite{Duan_PRL_2000,Simon_PRL_2000}
\begin{equation}
C > \sqrt{1-A-B+AB}.
\label{ent_criterion}
\end{equation}
In particular, in the case $A=B=S=2 N_s+1$, we have $C > C_c = 2N_s$, where $C_c$ represents the largest value of the correlations $C$ that can be obtained by classical means, i.e.\@ without entanglement. 
From the expressions for $C_q$ and $C_c$, we see that the two-mode squeezed vacuum is always entangled, but also that the correlation-enhancement enabled by entanglement ($C_q > C_c$) becomes less and less important as we increase the number of signal photons $N_s$.
We will see that this fact plays a fundamental role in understanding the quantum advantage in quantum illumination.
It is worth mentioning that the uncertainty principle \eqref{uncertainty} imposes a bound on the off diagonal terms of the covariance matrix \eqref{cov_TM}, which in the case of $A=B=S=2N_s+1$ reduces to $C \leq \sqrt{S^2 -1}=2N_s\sqrt{1 + 1/N_s}$.
Which means that for a fixed value of $S$, i.e.\@ a fixed number of photons $N_s$ in the signal, the two-mode squeezed vacuum achieves the largest possible values of the correlations $C$.

\begin{wrapfigure}{r}{0.5\textwidth}
\centering
\begin{tikzpicture}[scale=0.5]
\usetikzlibrary{arrows}
\coordinate (A1) at (0,1);
\coordinate (A2) at (1,1);
\coordinate (A3) at (1,4);
\coordinate (A4) at (0,4);
\coordinate (A5) at (2,2);
\coordinate (A6) at (2,5);
\coordinate (A7) at (1,5);
\coordinate (A8) at (1,2);
\coordinate (P1) at (-2.5,3);
\coordinate (P2) at (0.5,3);
\coordinate (S1) at (1.5,3);
\coordinate (S2) at (4,4.5);
\coordinate (I2) at  (4,1.5);
\draw[blue,->,line width=1pt]  (P1) -- node[above,xshift=-3] {$\omega_P,k_P$} (P2);
\draw[-triangle 90 cap, line width=6pt,blue,opacity=0.2] (-2.6,3) -- (0.6,3);
\draw[thick,fill=cyan, draw=cyan, fill opacity=0.4] (A1) -- (A2) -- (A3) -- (A4) -- cycle;
\draw[thick,fill=cyan, draw=cyan, fill opacity=0.4]  (A2) -- (A5) -- (A6) -- (A3)--cycle ;
\draw[thick,fill=cyan, draw=cyan, fill opacity=0.4]  (A3) -- (A6) -- (A7) -- (A4)--cycle ;
\draw[red,->,line width=1pt]  (S1) -- (S2) node[above] {$\omega_S,k_S$} ;
\draw[-triangle 90 cap, line width=6pt,red,opacity=0.2] (S1) -- (4.05,4.5*4.05/4);
\draw[red,->,line width=1pt]  (S1) -- (I2) node[below] {$\omega_I,k_I$};
\draw[-triangle 90 cap, line width=6pt,red,opacity=0.2] (S1) -- (4.05,1.45*4.05/4);
\fill[red] (S1) circle (0.1);
\draw [red] (4,3) ellipse (0.5 and 1.5);
\draw [red,line width=3pt,red,opacity=0.2] (4,3) ellipse (0.5 and 1.5);
\node at (9,4) {$\omega_P = \omega_S +\omega_I$};
\node at (9,2) {$k_P = k_S +k_I$};
\end{tikzpicture}
\caption{Pictorial representation of the non-linear process used to generate entangled photon pairs.}
\label{Fig:SPDC}
\end{wrapfigure}
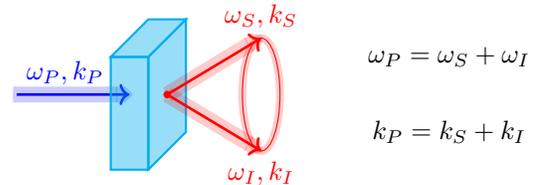
Let us conclude this section by saying that the two-mode squeezed vacuum is the entangled state which is most commonly generated in physics laboratories. 
It is normally produced using a non-linear process called spontaneous parametric down conversion (SPDC) in which a pump beam with frequency $\omega_P$ and wave vector $k_P$ is converted in correlated pairs of photons {\it signal} and {\it idler} modes characterized by the frequencies $\omega_{S/I}$ and the wave vectors $k_{S/I}$ (see Fig.\@ \ref{Fig:SPDC} ) \cite{agarwal_2012}.
For such a process to happen it is necessary to respect at the same time energy and momentum conservation. 
The latter is often referred to in this context as {\it phase matching condition}, which is verified only within a limited frequency range known as the {\it phase matching bandwidth}.
SPDC can be observed at optical frequencies in non-linear crystals, also known as optical parametric amplifiers (OPA) \cite{agarwal_2012}, as well as at microwave frequencies, where the analogue of an OPA is a superconducting circuit known as Josephson parametric amplifier (JPA) \cite{Flurin_PRL_2012}.

\subsection{Discriminating quantum states and probability distributions}
\label{sec:discrimination}
Let us now consider the problem of deciding between two hypotheses $H_0$ and $H_1$ based on some prior knowledge of the conditional probability densities $p_0({\bm R})$ and $p_1({\bm R})$ for the random variable ${\bm R}$ when one of the two hypotheses is true.
The problem of deciding between one of the two hypotheses is therefore equivalent to sample from two different probability distributions and discriminate them from the sampling results.
In the following, we will discuss the case when $p_0({\bm R})$ and $p_1({\bm R})$ are classical probability densities, as well as the one when they originate from the quantum states $\rho_0$ and $\rho_1$ corresponding to the two hypotheses.

Let us assume that ${\bf R}$ spans a real space $Z$ which we call the {\it decision space}.
To decide between the two hypotheses, let us divide the decision space $Z$ into two regions $Z_0$ and $Z_1$.
When ${\bf R} \in Z_0$ we decide that $H_0$ is true, and when ${\bf R} \in Z_1$, we decide otherwise.
How one constructs the regions $Z_0$ and $Z_1$ in order to minimize the error in choosing which hypothesis is true defines a {\it decision strategy} \cite{VanTrees}.
To better define a decision strategy, let us examine the four possible outcomes of a binary hypothesis test
\begin{enumerate}
\item decide that $H_0$ is true, when  $H_0$ is true,
\item decide that $H_1$ is true, when  $H_0$ is true,
\item decide that $H_1$ is true, when  $H_1$ is true,
\item decide that $H_0$ is true, when  $H_1$ is true,
\end{enumerate}
where cases 1.\@ and 3.\@ correspond to correct decisions, while cases 2.\@ and 4.\@ correspond to errors. 
Having in mind the radar problem, where $H_0$ corresponds to target absent and $H_1$ corresponds to target present, we will refer to case 2.\@ as a {\it false alarm} error, and to case 4.\@ as a {\it miss} error. 
Accordingly, we define the false alarm probability $P_F$, and the miss probability $P_M $ associated with the errors 2.\@ and 4.\@ respectively \cite{VanTrees}. In particular, the miss probability is related to the probability of correctly detecting the target $P_D$ by the relation $P_D = 1 - P_M$.

Given the above definitions, we can follow two different decision strategies: either we construct $Z_0$ and $Z_1$ in order to minimize the mean error probability $P_e = w_0 P_F + w_1P_M$, with $w_{0/1}$ the prior probabilities of $H_{0/1}$ to be true, or we fix a threshold for the false alarm probability $P_F$ and minimize the miss probability $P_M$ (or maximize the detection probability $P_D$) under this constraint.
The first of these two strategies is often referred to as the Bayesian approach, while the second one is known as the Neyman--Pearson approach \cite{VanTrees}. 

In the next sections, we will describe these two decision strategies following both from a classical and a quantum point of view.

\subsubsection{Bayesian approach: the Chernoff bounds}
\label{sec:Bayes}

{\bf Classical Chernoff bounds} 

It can be proved that an optimal test according to the Bayesian approach can be cast in the form of a likelihood test \cite{VanTrees}
\begin{equation}
\Lambda({\bm R}) = \frac{p_1({\bm R})}{p_0({\bm R})} \underset{H_0}{\overset{H_1}{\gtrless}} \gamma,
\label{Bayes_decision_rule}
\end{equation}
where $\Lambda({\bm R})$ is known as the {\it likelihood ratio} and $\gamma = w_0/w_1$ is the decision threshold.\footnote{In expressing the probability of error, we assumed that all kinds of error come with the same cost. In a more general framework, it is possible to associate different costs to the false alarm and miss errors, and even to introduce costs for the correct decisions \cite{VanTrees}. The values of these costs will modify the form of the decision threshold $\gamma$. However, in practice, it is very difficult to assign costs and a priori probabilities, and the Bayesian strategy is often used under the assumptions that both hypotheses are equally probable, and all errors have the same impact. 
A more practical way to  deal with different impacts of different kinds of errors is provided  by the Neyman--Pearson decision strategy that we will describe in Sec.\@ \ref{sec:Neyman--Pearson}.}
Accordingly, the regions $Z_0$ and $Z_1$ of the decision space are defined by the conditions 
\begin{equation}
Z_0 = \left\{ {\bf R}\; {\rm s.t.}\; \frac{p_0({\bf R})}{p_1({\bf R})} > \frac{w_1}{w_0} \right \} \; {\rm and} \; Z_1 = \left\{ {\bf R}\; {\rm s.t.}\; \frac{p_1({\bf R})}{p_0({\bf R})} > \frac{w_0}{w_1} \right \}  
\label{Bayes_Z}
\end{equation}
It is interesting to notice that when both hypotheses are equally probable ($w_0 = w_1 =1/2$), we have $\gamma =1$. 

In addition to know which test one needs to perform, it is crucial to estimate which error probability is associated with a given test as defined by $Z_0$ and $Z_1$. 
Let us therefore consider the mean error probability
\begin{align}
P_e = w_0 P_F + w_1P_M &=  w_0 \int_{Z_1} p_0({\bm R}) {\rm d}{\bm R} + w_1 \int_{Z_0} p_1({\bm R}) {\rm d}{\bm R} \nonumber\\
& = \int_Z \min \left[w_0 p_0({\bm R}), w_1 p_1({\bm R})\right]{\rm d}{\bm R},
\label{P_e_bayes}
\end{align}
where we used that by definition the Bayesian decision strategy minimize the mean error probability.
We can now obtain a bound on the mean error probability by using the fact that for every pair of positive numbers $a$ and $b$, $\min (a,b) \leq a^s b^{1-s}$, with $0 \leq s \leq 1$
\begin{equation}
P_e \leq \min_{0 \leq s \leq 1} w_0^s w_1^{1-s}\int p_0^s({\bm R}) p_1^{1-s}({\bm R}){\rm d}{\bm R}.
\end{equation}
In the following, we will mostly focus on the case $w_0 = w_1 =1/2$, corresponding to the case where no prior information on the target presence/absence is available. Under this assumption, we can write 
\begin{equation}
P_e \leq \frac{1}{2} e^{- M \xi_{C}}  = \frac{1}{2} \min_{0 \leq s \leq 1}\int p_0^s({\bm R}) p_1^{1-s}({\bm R}){\rm d}{\bm R},
\label{Chernoff_cl}
\end{equation}
where we have introduced the error probability exponent $\xi_C$
and the number of independent, identically-distributed samples $M$ used for the hypothesis testing.
The bound on the error probability in Eq.\@ \eqref{Chernoff_cl} is known as the {\it Chernoff bound}, and it has the remarkable property of being asymptotically tight \cite{Nielsen_2011}, i.e.\@
\begin{equation}
\xi_{C}  = -\lim_{M \to \infty} \frac{\log P_e}{M}.
\end{equation}
The minimization over $s$ is often not a trivial task. 
It is therefore useful to introduce the simpler {\it Bhattacharyya bound},
\begin{equation}
P_e \leq \frac{1}{2} e^{- M \xi_{C}}  \leq \frac{1}{2} e^{- M \xi_{B}} =  \frac{1}{2} \int \sqrt{ p_0({\bm R}) p_1({\bm R})}\rm{d}{\bm R},
\label{classical_Bhattacharyya}
\end{equation}
which however is not exponentially tight. 
In our discussion of quantum illumination, we will only deal with Gaussian probability distributions for which the Chernoff and the Bhattacharyya bounds are often easy to calculate.

{\bf Quantum Chernoff bounds} 

From the quantum mechanical point of view, we have the density operators $\rho_{0/1}$ associated with the quantum state of the system under hypotheses $H_{0/1}$, and the measurement operators $E_{0/1}$ corresponding to our decision strategy.
Accordingly, as in the classical case before, we chose the regions $Z_{0/1}$ in the decision space in order to minimize the mean error probability, we will now choose the operators $E_{0/1}$ that minimize 
\begin{equation}
P_e = w_0 {\rm tr} (E_1 \rho_0) + w_1 {\rm tr} (E_0 \rho_1).
\label{Pe}
\end{equation}
To do this minimization, we use\footnote{If we think about the analogy between the operators $E_{0/1}$ and the regions of the decision space $Z_{0/1}$ used in the classical description, the condition $E_0 + E_1 = \mathds{1}$ corresponds to the fact that the union of the two regions $Z_0$ and $Z_1$ corresponds to the full space $Z = Z_0 \bigcup Z_1$.}
$E_0 = \mathds{1} - E_1$, and we rewrite Eq.\@ \eqref{Pe} as
\begin{equation}
P_e  = w_1 - {\rm tr}\left[ E_1 \left(w_1\rho_1 - w_0\rho_0 \right)\right].
\end{equation}
If we now consider that the trace of an operator is equal to the sum of its eigenvalues, we can minimize the above expression by taking $E_1$ such that $E_1 (w_1\rho_1 - w_0\rho_0) = (w_1\rho_1 - w_0\rho_0)_+$,
where we have introduced the notation $A_+$ for the positive part of the operator $A$, namely $A$ 
restricted to the subspace corresponding to its positive eigenvectors.
The latter can be expressed in term of the operator absolute value $|A| = (A^*A)^{1/2}$ as $A_+ = (|A|+A)/2$, therefore we can write 
\begin{equation}
P_e = \frac{1-{\rm tr} |w_1\rho_1 - w_0\rho_0|}{2}.
\label{Helstrom_Bound}
\end{equation}
We therefore have that the projector on the positive part of $w_1\rho_1 - w_0\rho_0$ is the quantum analogue of the classical Bayesian decision rule \eqref{Bayes_decision_rule}\footnote{The reader should also notice that the classical likelihood-ratio test can be rewritten as: decide $H_1 (H_0)$ if $w_1 p_1({\bf R}) - w_0p_0({\bf R})$ is positive (negative).}, in the sense that it achieves the minimal mean error probability given by Eq.\@ \eqref{Helstrom_Bound}, which is known as the {\it Helstrom bound} \cite{Helstrom}.

We will be interested in evaluating the mean error probability when we have access to $M$ copies of the system under study, in this case the Helstrom bound take the form 
\begin{equation}
P_e = \frac{1-{\rm tr}|w_1\rho_1^{\otimes M} - w_0\rho_0^{\otimes M}|}{2}, 
\end{equation}
which is however often very difficult to calculate in practice. 
Luckily, quantum versions of the Chernoff and Bhattacharyya bounds exists, and their explicit forms bear a striking resemblance with their classical analogues \cite{Audenaert_PRL_2007,Audenaert2008}
\begin{equation}
P_e \leq \frac{1}{2} e^{- M \xi_{QC}}  \leq \frac{1}{2} e^{- M \xi_{QB}},
\label{quantum_exp}
\end{equation}
with 
\begin{equation}
\xi_{QC} = -\log \left[\min_{0 \leq s \leq 1} {\rm tr}\left( \rho_0^s\rho_1^{1-s} \right) \right],
\label{Quantum_chernoff}
\end{equation}
and
\begin{equation}
\xi_{QB} = -\log \left[\rm{tr}\left( \sqrt{\rho_0}\sqrt{\rho_1} \right) \right],
\label{Quantum_Bhattacharyya}
\end{equation}
where for simplicity, we restricted ourselves to the case $w_0 = w_1 = 1/2$.
Interestingly enough, the Bhattacharyya bound can be related to a lower bound for the mean error probability \cite{Weedbrook_Rev_Mod_Phys_2012}
\begin{equation}
P_e \geq \frac{1}{2}\left(1 - \sqrt{1 - e^{-M\xi_{QB}}} \right).
\label{lower_bound}
\end{equation}

As in the classical case, the quantum Chernoff bound is asymptotically tight, while the Bhattacharyya bound (and the lower bound derived from it) is not as tight, but is much simpler to calculate \cite{Audenaert_PRL_2007,Audenaert2008}.
For a Gaussian state with known mean vector and covariance matrix, it is always possible to obtain an analytical expression for $\rm{tr}\left( \rho_0^s\rho_1^{1-s} \right)$ \cite{Pirandola_PRA_2008}. 
However, the optimization over $s$ in Eq.\@ \eqref{Quantum_chernoff} cannot always be carried out analytically. 
When this is the case, we will resort to the Bhattacharyya bound in Eq.\@ \eqref{Quantum_Bhattacharyya}.

\subsubsection{Neyman--Pearson approach: the receiving operating characteristic}
\label{sec:Neyman--Pearson}

{\bf Classical Neyman--Pearson approach} 

In the previous section, we described the Bayesian decision strategy, which aims to minimize the mean error probability.
However, this strategy is not the most appropriate for problems in which different types of error have not the same importance, as it is the case in target detection.
In fact, especially if we think about military applications, the damage produced by missing an enemy plane can be much larger than that associated with a false alarm.
In such contests, it is more convenient to establish a value of the false alarm probability that one can tolerate, $P_F =\beta$, and then minimize the miss probability $P_M$, or equivalently maximize the detection probability $P_D = 1 -P_M$.

This is a constrained optimization problem that can be solved by using Lagrange multipliers, namely by minimizing the function \cite{VanTrees}
\begin{equation}
F = P_M +\lambda\left[P_F - \beta \right] = \int_{Z_0} p_1({\bm R}){\rm d}{\bm R} + \lambda\left[\int_{Z_1} p_0({\bm R}){\rm d}{\bm R} - \beta\right].
\end{equation}
Since $\int_{Z_0} p_0({\bm R}){\rm d}{\bm R} + \int_{Z_1} p_0({\bm R}){\rm d}{\bm R} =1$, 
we can rewrite this function as 
\begin{equation}
F=\lambda(1-\beta) + \int_{Z_0} \left[p_1({\bm R}) -\lambda  p_0({\bm R})\right]{\rm d}{\bm R},
\label{NPintegral}
\end{equation}
and see that we can minimize $F$ by assigning a point to the decision region $Z_0$ whenever the argument of the integral in Eq.\@ \eqref{NPintegral} is negative. 
This corresponds to a likelihood test \cite{VanTrees}
\begin{equation}
\Lambda({\bm R}) = \frac{p_1({\bm R})}{p_0({\bm R})} \underset{H_0}{\overset{H_1}{\gtrless}} \lambda,
\label{NP_decision_rule}
\end{equation}
where the value of the threshold $\lambda$ is determined by the constraint condition
\begin{equation}
P_F = \int_\lambda^\infty p_0(\Lambda)\rm{d}\Lambda = \beta,
\label{NP_threshold}
\end{equation}
where we have denoted with $p_0 (\Lambda)$ the conditional probability density of the likelihood ratio $\Lambda$ under the assumption that the hypothesis $H_0$ is true.
From Eq.\@ \eqref{NP_threshold}, we notice that, as opposed to the Bayesian decision threshold $\gamma$, to determine the Neyman--Pearson threshold $\lambda$ one does not need to make any assumption on the a priori probabilities of the two hypotheses $w_0$ and $w_1$.

The performances of a Neyman--Pearson test are usually evaluated by plotting the detection probability 
\begin{equation}
P_D = 1 - P_M = \int_\lambda^\infty p_1 (\Lambda)\rm{d}\Lambda,
\label{NP_PD}
\end{equation}
as a function of the false alarm probability $P_F$.
The curve $P_D(P_F)$ is known as the receiver operating characteristic (ROC), and can be used to determine the optimal operating point of a given detector.

Analytical results for the integrals in Eqs.\@ \eqref{NP_threshold} and \eqref{NP_PD} are available when the probability densities $p_{0/1}({\bm R})$ are Gaussian functions, as it will be the case in quantum illumination theory (see Sec.\@ \ref{sec:gaussianQI}).
These results are exact in the case of two Gaussians with different means but the same variance, as well as in the opposite scenario of two Gaussians of same mean and different variances \cite{VanTrees}.
In the general case of Gaussians with different means a different variances an approximation known as the {\it extended Van Trees receiving characteristic approximation} is available \cite{Shapiro_IEEE_1999}.

{\bf Quantum Neyman--Pearson approach} 

Let us now discuss how the decision strategy discussed above translate to the the quantum case.
As in Sec.\@~\ref{sec:Bayes}, we consider the measurement operators $E_0$ and $E_1$ such that $E_0 + E_1 = \mathds{1}$, which define the false alarm, miss and detection probabilities as
\begin{subequations}
\begin{align}
P_F &= \rm{tr} \left( E_1 \rho_0\right), \label{PF}\\
P_M &= \rm{tr} \left( E_0 \rho_1\right), \label{PM}\\
P_D &= 1-P_M = \rm{tr} \left[ \left(\mathds{1} - E_0\right) \rho_1\right] = \rm{tr} \left( E_1 \rho_1\right). \label{PD}
\end{align}
\end{subequations}
We now want to find the operator $E_1$ that maximize Eq.\@ \eqref{PD} (or minimize Eq.\@ \eqref{PM}) for a fixed value of  the false alarm probability $P_F = \beta$.
By repeating the procedure that led to Eq.\@ \eqref{NPintegral} with the probability distributions replaced by density matrices and the integrals replaced by traces, we find that $E_1$ must be the projector on the positive part of $\left(\rho_1 -\lambda\rho_0 \right)$, with $\lambda$ determined by the condition $P_F = \beta$ via Eq.\@ \eqref{PF} \cite{Helstrom} \footnote{It should be noted that it is not always possible to saturate the condition $P_F = \rm{tr} [ E_1 \rho_0] = \beta$, with $E_1$ the projector $\Pi_+$ on $\left(\rho_1 -\lambda\rho_0 \right)_+$ by simply tuning $\lambda$. 
When this is not possible, one can define $E_1 = \Pi_+ + \xi \Pi_0,$ where $\Pi_0$ is the projector on the the subspace spanned by the eigenstates corresponding to vanishing eigenvalues of $\left(\rho_1 -\lambda\rho_0 \right)$ , i.e. the zero eigenspace.
In this case, $\lambda$ is tuned to reach the largest value of $\rm{tr} [ \Pi_0 \rho_0] \leq \beta$, and $\xi$ is then chosen to reach the equality $P_F = \beta$ \cite{Helstrom}. This is particularly relevant when dealing with states $\rho_{0/1}$ with discrete spectra (e.g.qubits). On the other hand, when dealing with states with a continuous spectrum the zero eigenvalues subspace is a set of measure zero \cite{Helstrom}. 
In the following, we will apply this formula in the latter case, and we will ignore the mathematical complications coming from the zero eigenspaces. 
}.

Following  this strategy one can determine a quantum version of the ROC. 
To illustrate this idea, let us now consider the binary discrimination between two pure states $\rho_0 = \ket{\psi_0}\bra{\psi_0}$ and $\rho_1 = \ket{\psi_1}\bra{\psi_1}$, which are in general non-orthogonal $\braket{\psi_1|\psi_0} = \nu$.
To find the projector on the positive part of $\left(\rho_1 -\lambda\rho_0 \right)$, we thus have to solve the eigenvalue equation
\begin{equation}
\left(\ket{\psi_1}\bra{\psi_1} - \lambda \ket{\psi_0}\bra{\psi_0} \right)\ket{\eta} =\eta \ket{\eta}.
\end{equation}
Multiplying from the left by $\bra{\psi_0}$ and $\bra{\psi_1}$, we obtain the system of equations
\begin{equation}
\begin{cases}
\nu^*\braket{\psi_1|\eta}- (\eta +\lambda)\braket{\psi_0|\eta}=0\\
(1-\eta)\braket{\psi_1|\eta} - \nu\braket{\psi_0|\eta}=0
\end{cases},
\label{NP_sys}
\end{equation}
which leads to the solutions
\begin{equation}
\eta_{1/0} = \frac{1}{2}\left(1 - \lambda \right) \pm R, \quad {\rm with} \; R = \sqrt{(1-\lambda)^2/4 +\lambda h}, \quad {\rm and}\; h = 1-|\nu|^2.
\label{NP_eig}
\end{equation}
It is easy to verify that $\eta_1>0$ and $\eta_0 <0$. Accordingly, the Neyman--Pearson criterion tells us to define $E_{0/1} = \ket{\eta_{0/1}}\bra{\eta_{0/1}}$.
Substituting into Eqs. \eqref{PF} and \eqref{PD}, we have $P_F = |\braket{\psi_0|\eta_1}|^2$ and $ P_D = |\braket{\psi_1|\eta_1}|^2$, where $\braket{\psi_{0/1}|\eta_1}$ can be determined by substituting Eq.\@ \eqref{NP_eig} into the system \eqref{NP_sys}.
Finally, one obtains the false alarm and detection probabilities as functions of $\lambda$,
\begin{subequations}
\begin{align}
P_F &= (\eta_1 - h)/2R, \\
P_D &= (\eta_1 +\lambda h)/2R.
\end{align}
\end{subequations}
Eliminating $\lambda$, we get an analytical expression for the ROC  \cite{Helstrom}
\begin{equation}
P_D = 
\begin{cases}
\left(\sqrt{P_F(1-h)}+ \sqrt{(1-P_F)h}\right)^2 & \text{when } 0 \leq P_F \leq 1-h\\
1&\text{when } 1-h \leq P_F \leq 1
\end{cases}
\label{ROC_pure}
\end{equation}
This equation is valid only for pure states. In quantum illumination, we will be interested in detecting a target located in a region with a strong thermal background. Because of this thermal background, all quantum states we will consider will be statistical mixtures. A general quantum framework to discriminate between two arbitrary (eventually mixed) Gaussian states following the Neyman--Pearson approach actually exists \cite{PhysRevLett.119.120501}.
However, in the following, to keep things simple, we will focus on specific measurement strategies.
In this context, the quantum state discrimination problem is mapped into the discrimination between two classical probability distributions, that we can tackle with the classical theory described above.
Moreover, we will see that under certain conditions the output of the ultimate receiver for quantum illumination can be approximated with a pure state, and therefore we will use Eq.\@ \eqref{ROC_pure} in Sec.\@ \ref{sec:performances} to roughly estimate its ROC.

\section{Quantum illumination theory}
This part represents the core of this review. 
Here, we will use all the concepts introduced above to describe in quite some details the theory of quantum illumination for target detection. 
We will start by presenting the quantum illumination protocol based on Gaussian states introduced by Tan \emph{et al.\@} \cite{Tan_PRL_2008}.
We will then discuss different detection strategies that allow to exploit the quantum advantage enabled by this protocol and we will quantify its performances.
Finally, we will comment on various critical points and limitations of this protocol that severely limits its practical usefulness.

\subsection{Gaussian quantum illumination}
\label{sec:gaussianQI}
Let us now imagine that we transmit a signal to a region that contains a strong thermal background and may or may not contain a weakly reflecting target, and by measuring the returned light we want to discriminate between the hypothesis $H_0$ (target absent) and the hypothesis $H_1$ (target present).
When the hypothesis $H_0$ is true, the signal coming back from the target region is defined by the annihilation operator $\hat{a}_R = \hat{a}_B$, where the mode $\hat{a}_B$ is in a thermal state with average photon number $N_B \gg 1$.
We focus on the case of strong thermal background ($N_B \gg 1$), since it is the most relevant for microwave radar applications. Moreover, it has been proved that quantum light gives no benefit when the background is absent or weak \cite{Nair:2011}.

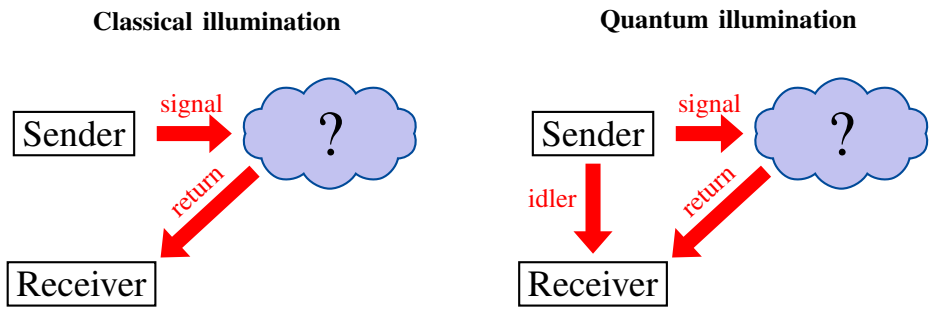
\begin{figure}[htb]
\centering
\begin{tikzpicture}
\tikzset{arw/.style={>={Triangle[length=3mm,width=5mm]},line width=2mm,draw=red}}
\path[use as bounding box] (0,3.5) -- (11.5,3.5) -- (11.5,7.5) -- (0, 7.5) -- cycle;    
\node at (2.,7.5) {\bf Classical illumination};
\node at (3.5,6) [cloud, draw = MyBlue, thick, fill=gray!40!blue!30!, cloud puffs=10, cloud puff arc=120, aspect=2, inner ysep=0.3 em] {\Huge ?};
\node at (0.1,6)  [rectangle, draw, thick] {\Large Sender};
\draw[arw,->] (1.2,6) --  node[above,red,xshift=-1] {signal} (2.2,6);
\node at (0.2,4) [rectangle, draw, thick] {\Large Receiver};
\draw[arw,->] (2.5,5.5) --  node[pos=0.5,sloped,above,red,xshift=3] {return} (1.25,4.35);
\node at (2.+6.8,7.5) {\bf Quantum illumination};
\node at (3.5+6.8,6) [cloud, draw = MyBlue, thick, fill=gray!40!blue!30!, cloud puffs=10, cloud puff arc=120, aspect=2, inner ysep=0.3 em] {\Huge ?};
\node at (0.2+6.8,6)  [rectangle, draw, thick] {\Large Sender};
\draw[arw,->] (1.3+6.8,6) --  node[above,red] {signal} (2.2+6.8,6);
\node at (0.2+6.8,4) [rectangle, draw, thick] {\Large Receiver};
\draw[arw,->] (2.5+6.8,5.5) --  node[pos=0.5,sloped,above,red,xshift=3] {return} (1.25+6.8,4.35);
\draw[arw,->] (0.2+6.8,5.6) --  node[left,red,yshift=4] {idler} (0.2+6.8,4.4);
\end{tikzpicture}
\caption{Pictorial representation of a classical target detection scheme (on the left) and a quantum one based on entanglement (on the right).}
\label{Fig:QI}
\end{figure}

When hypothesis $H_1$ is true, the return-mode's annihilation operator is given by
\begin{equation}
\hat{a}_R = \sqrt{\kappa}\hat{a}_S +\sqrt{1-\kappa} \hat{a}_B,
\end{equation}
where $\kappa \ll 1$ is the radar-to-target round-trip transmissivity, while the thermal state defined by $\hat{a}_B$ now contains $N_B/(1-\kappa)$ photons, such that the total number of received background photons is the same under both hypotheses \footnote{It should be noted that for $\kappa \ll 1$ the difference between $N_B$ and $N_B/(1-\kappa)$ is quite small. On the other hand, this assumption is crucial to ensure that there is no passive signature of target presence that could be sensed without transmission.}.
We will now determine how to discriminate between these two hypotheses in the two scenarios illustrated in Fig.\@~\ref{Fig:QI}. 

In the classical illumination scenario, we send the coherent state $\ket{\sqrt{N_s}}$ to the target region and we measure the returned light, which is either in a strong thermal state ($H_0$ is true) or in a combination of a very weak coherent state and a much stronger thermal state ($H_1$ is true).
Under both hypothesis, the returned state is a Gaussian state of the form \eqref{Gaussian_W} with mean vectors $\bar{\bm x}_0 =(0,0)$ (under $H_0$) and $\bar{\bm x}_1 =(2\sqrt{\kappa N_s},0)$ (under $H_1$).
The covariance matrix of the return mode is the same under both hypothesis and is given by ${\bm V}_0 = {\bm V}_1 = B\mathds{1}_2 $, with $B = 2N_B+1$.
In this scenario, discriminating between the two hypotheses $H_0$ and $H_1$ corresponds to distinguishing between a very broad Gaussian distribution and an identical one which is slightly shifted with respect to the first one.

In the quantum illumination scenario, we produce a two-mode squeezed vacuum state, and we retain the idler at the sender, while we transmit the signal to the target region.
In order to decide which one of the two hypotheses is true, we perform this time a {\it joint measurement} on the return signal and the retained idler.
By joint measurement, we mean a detection scheme that addresses the signal and the idler together as a single quantum system and is therefore able to witness their correlations. 
As we will discuss in details in the following, measuring individually the signal and the idler is not enough to reconstruct their correlations in post-processing. 

Under both hypothesis, the quantum state of the return and idler modes is Gaussian with mean vector $\bar{\bm x}_0 = \bar{\bm x}_1 =(0,0,0,0)$ and covariance matrices
\begin{equation}
{\bm V}_0 =
\begin{pmatrix}
B & 0 & 0 & 0\\
0 & B & 0 & 0\\
0 & 0 & S & 0\\
0 & 0 & 0 & S
\end{pmatrix}, \quad
{\bm V}_1 =
\begin{pmatrix}
A & 0 & \sqrt{\kappa}C_q & 0\\
0 & A & 0 & -\sqrt{\kappa}C_q &\\
\sqrt{\kappa}C_q & 0 & S & 0\\
0 & -\sqrt{\kappa}C_q & 0 & S
\end{pmatrix},
\label{QIcov}
\end{equation}
with $S = 2N_s +1$, $A = 2\kappa N_s +B$, which for $\kappa N_s \ll 1$, corresponds to $A \sim B$. 

Let us notice that both these covariance matrices correspond to separable states.
In fact, when the target is absent, ${\bm V}_0$ in Eq.\@ \eqref{QIcov} is diagonal (does not contain any correlation term). 
On the other hand, by comparing ${\bm V}_1$ in Eq.\@ \eqref{QIcov} with Eq.\@ \eqref{ent_criterion}, we see that even when the target is present, the return mode is not entangled whenever $\kappa < N_B$, which is always the case when there is more than one photon in the thermal background.
In other words, under both hypotheses the received quantum states could be described as classical Gaussian noise processes.
However, we will see that even if the initial entanglement is completely destroyed by losses and noise in the channel, the surviving correlations between the return mode and the retained idler are enough to provide a quantum advantage in the weak signal regime.

\begin{SCfigure}[][hbt]
\centering
\includegraphics[width=0.35\textwidth]{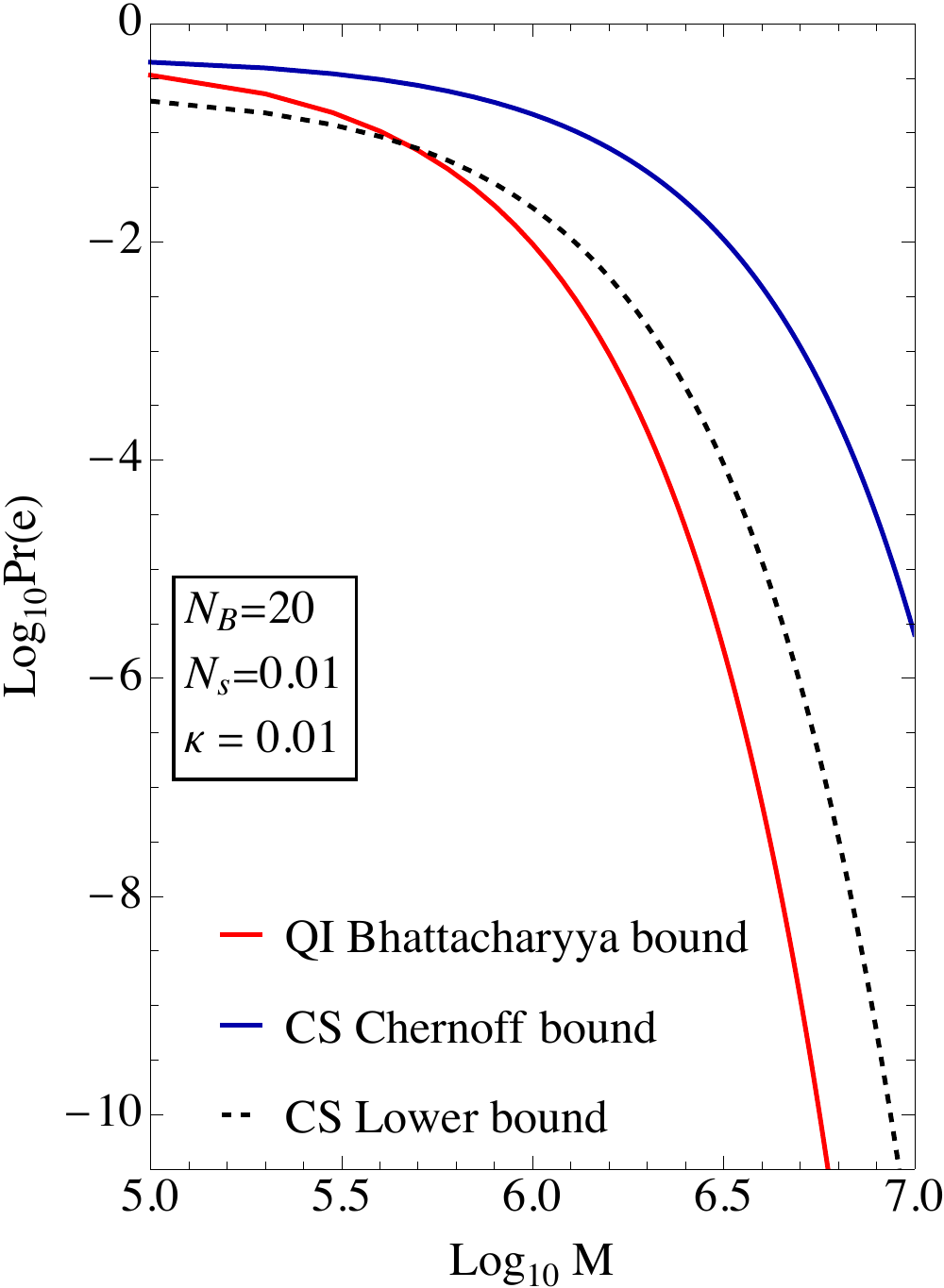}
\includegraphics[width=0.35\textwidth]{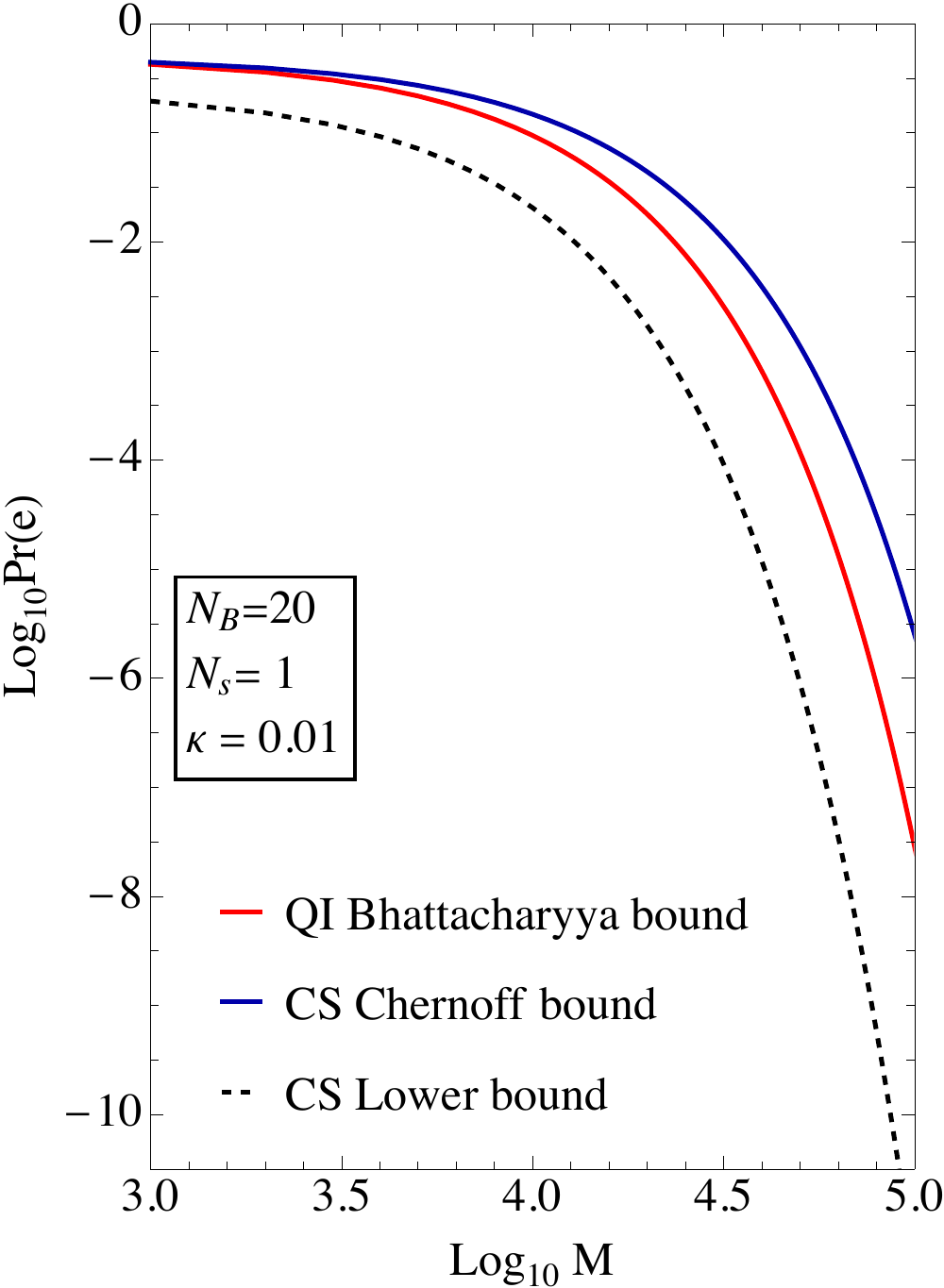}
\caption{Upper bounds for the mean error probability for classical (blue) and quantum (red) illumination as a function of the number of copies $M$ of the state used to interrogate the target region. 
We also included the lower bound (black dashed) for classical illumination.
On the left, we see that in $N_s \ll1$ case there is a clear regime where the QI illumination upper bound falls below the coherent states lower bound. 
On the other hand, on the right hand side, we see that already for $N_s =1$ the quantum illumination upper bound falls between the bounds for classical illumination.
}
\label{Fig:QI_Gauss}
\end{SCfigure}

In the following, we compare the performance of classical and quantum illumination by evaluating the mean error probabilities, under the assumption that both type of error has the same importance ($w_0 = w_1 =1/2$), when $M$ copies of the state are sent to the target region. 
The number of copies $M$ should be understood here as the number of modes, as coming from a Fourier-series decomposition of an electromagnetic pulse of duration $T$ with a frequency bandwidth $W$, namely as the time-bandwidth product $M = TW$.

The time-bandwidth product $M = TW$ will play a fundamental role in the discussion of the practical limitations of quantum illumination (Sec.~\ref{Sec:criticalities}). Therefore, some clarifications are in order. 
For quantum illumination, the bandwidth $W$ is determined by the phase matching bandwidth of the SPDC process used to generate the entangled signal idler pulses (see Sec.~\ref{sec:gaussian}).
Radar engineers should note that the phase matching bandwidth of SPDC simply tells us over which range of frequencies we can produce entangled photons, but it does not correspond to a known temporal modulation of the frequency content of the signal. 
In this sense, the high time-bandwidth product signal necessary for quantum illumination is very different from the deterministic high time-bandwidth product signal used in classical pulse compression radars.
The way quantum illumination uses its high time-bandwidth product to detect phase-sensitive correlations between the return signal and the retained idler is more similar to a noise radar\footnote{An even closer analogy can be drawn in the case of quantum-correlated noise radars which use SPDC sources and pre-amplified heterodyne detection (see Sec.~\ref{sec:receivers}). We will not discuss the theory of these quantum radars in detail since they can never overcome their classical counterparts \cite{ShapiroReview2019}. But we will present the limitations of their experimental implementations \cite{luong2019receiver, Barzanjeh_arxiv_2019} in Sec.~\ref{Sec:experiments}.}. 
The latter obtains its pulse compression gain by correlating the signal returning from the interrogation region with a retained noise reference perfectly correlated with the transmitted signal.
However, we should be careful in using this analogy with classical noise radars to obtain a pulse-compression interpretation of quantum illumination. 
In fact, as we will discuss in detail in Sec.~\ref{Sec:criticalities}, quantum illumination cannot make a cross-correlation measurement across a span of possible target-range delays without sacrificing its quantum advantage.

After these important clarifications, we are now ready to evaluate the mean error probabilities for classical and quantum illumination.
In the coherent state case, we can calculate the Chernoff bound \eqref{Quantum_chernoff} (which happens to coincide with the Bhattacharyya one) \cite{Tan_PRL_2008} which is given by
\begin{equation}
P_e^{CS} \leq \frac{1}{2}e^{-M\kappa N_s(\sqrt{N_B} - \sqrt{N_B+1})^2} \approx \frac{1}{2}e^{-M\kappa N_s/4N_B} = e^{-M\xi_C},
\label{CS_Pe}
\end{equation}
where the approximation is valid whenever $N_B \gg 1$.
Given that the Chernoff bound in \eqref{CS_Pe} coincides with the Bhattacharyya one, we can use it to compute the lower bound \eqref{lower_bound}, which gives
\begin{equation}
P_e^{CS} \geq \frac{1}{2}\left(1 - \sqrt{1 - e^{-M\kappa N_s(\sqrt{N_B} - \sqrt{N_B+1})^2}} \right) \approx \frac{1}{4}e^{-M\kappa N_s/2N_B}.
\label{lower_bound_CS}
\end{equation}
Using convexity arguments, it can be proved that the lower bound \eqref{lower_bound_CS} is actually a lower bound for every classical state, i.e.\@ a two mode state with correlations $C\leq C_c$ (see Eq.\@ \eqref{ent_criterion}) \cite{Tan_PRL_2008}. 

For the quantum illumination case, we can obtain an analytical expression for the Bhattacharyya bound, which is however too long and cumbersome to be presented here.
A simplified expression can be obtained in the limit $\kappa \ll 1$, $N_s \ll 1$ and $N_B\gg 1$ \cite{Tan_PRL_2008}
\begin{equation}
P_e^{QI} \leq \frac{1}{2}e^{-M\xi_{QI}} \approx \frac{1}{2}e^{-M\kappa N_s/N_B}.
\label{QI_Pe}
\end{equation}
Comparing Eqs.\@ \eqref{CS_Pe} and \eqref{QI_Pe}, we see that in the low-brightness ($N_S \ll 1$) limit, despite entanglement being lost, quantum illumination provides a $6$ dB  enhancement of the error exponent. Moreover, in this regime low-brightness regime, for large values of $M$, the upper bound for quantum illumination falls below the upper bound for coherent state illumination (see left panel in Fig.\@ \ref{Fig:QI_Gauss})showing a clear quantum advantage.

However, the full expression for $P_e^{QI}$ tells us that this quantum advantage becomes less and less important when increasing $N_s$ (right panel in Fig.\@  \ref{Fig:QI_Gauss}), and it ultimately disappears when $N_s \gg 1$.
This can be understood if we consider that the quantum advantage originates from the fact that the two-mode squeezed vacuum has stronger phase-sensitive correlations than any possible classical state, but that, as discussed in Sec.\@ \ref{sec:gaussian}, the difference between these correlations and those allowed by classical means becomes less and less important when the photon number is increased.

\subsection{Practical receivers for quantum illumination}
\label{sec:receivers}
In the previous section, we calculated bounds on the mean error probability for classical and quantum illumination. 
However, we left the question of how to achieve these bounds open. 
In fact, even though in Sec.\@ \ref{sec:Bayes} we derived the measurement operator that saturates the Helstrom bound \eqref{Helstrom_Bound}, and we argued that this is the analogue of the Bayesian decision rule \eqref{Bayes_decision_rule}, there is a substantial difference between the two.
Namely, that Eq.\@ \eqref{Bayes_decision_rule} can be easily applied to most experimental results, while the practical implementation of the projector on $(w_1\rho_1 -w_0\rho_0)_+$ is often impossible.
Here, we will address how to witness the quantum advantage offered by quantum illumination while using standard quantum optics detection schemes. 

There are substantially three kinds of measurement that can be easily  performed in quantum optics \cite{Weedbrook_Rev_Mod_Phys_2012}: 
\begin{enumerate}
\item {\bf Homodyne detection} is a measurement of the quadrature operator $\hat{q}$ (or $\hat{p}$) associated with a given mode, whose outcome is a continuous variable $q$ (or $p$) which follows the probability distribution $P(q)$ (or $P(p)$). In the particular case of a Gaussian state, these probability distributions are Gaussian too  \cite{Weedbrook_Rev_Mod_Phys_2012}.
Experimentally, homodyne detection is realized by mixing the quantum mode to be measured with a continuous-wave intense local oscillator, with central frequency matching that of the radiation to be measured, on a $50{:}50$ beam splitter, and by measuring the intensity of the output modes. 
The difference of these intensity is proportional to the quadrature $q$, while the quadrature $p$ can be measured by applying a $\pi/2$ phase shift to the local oscillator \cite{agarwal_2012}.
\item {\bf Heterodyne detection} is practically implemented by mixing the signal of interest with a continuous-wave, intense local oscillator whose central frequency is shifted with respect to that of the radiation to be measured.
The observable describing heterodyne detection can be proved to be equivalent to the one describing the mixing of the quantum mode of interest with a vacuum mode on a $50{:}50$ beam splitter, and then measuring via homodyne  detections the quadratures $q$ and $p$ of the two the output modes.\footnote{This procedure, known as {\it dual homodyne} is far more common at optical frequencies and, because of the equivalence discussed above, it is often referred to as heterodyne detection in the quantum optics literature.} 
Accordingly, heterodyne allows to access both field quadratures at the same time.
However, we cannot escape the fact that $\hat{p}$ and $\hat{q}$ do not commute, and therefore by measuring both quadratures at the same time, we incur in additional noise as imposed by the uncertainty relation  \cite{Weedbrook_Rev_Mod_Phys_2012}.
\item {\bf Photon counting} consists in measuring the number of photons in a certain mode, which after averaging over several measurements, gives $\langle\hat{n}\rangle = \langle \hat{a}^\dagger\hat{a}\rangle$.
\end{enumerate}

For a coherent state transmitter, we can show that homodyne detection allows to saturate the Chernoff bound \eqref{CS_Pe} \cite{Guha_PRA_2009}.
In fact, if we homodyne the return mode in the classical illumination protocol described in section \ref{sec:gaussianQI}, the probability distributions $p_{q_k}(Q_k|H_{0/1})$ of the measured quadrature $q_k$ is a Gaussian with variance $2N_B +1$ and mean $\bar{q}_k = 0$ and $\bar{q}_k = 2\sqrt{\kappa N_s}$ under hypotheses $H_0$ and $H_1$ respectively.
Here $k= 1, \cdots, M$ labels the different measurements on $M$ copies of the return mode.
These are now classical probability distributions, and we can apply classical decision theory do discriminate between hypotheses $H_0$ and $H_1$.
In particular, the Bayesian decision rule \eqref{Bayes_decision_rule} corresponds to evaluate the mean $Q_k$ of each of the $M$ homodyne measurements, to calculate $q = q_1 + \cdots+  q_M$, and decide in favour of $H_0$ whenever $Q < (M\sqrt{\kappa N_s})$, and in favour of $H_1$ otherwise \cite{VanTrees}.
Using the above definitions for $p_{q}(Q|H_{0/1})$, and the fact that $Q$ is also a Gaussian random variable we can rewrite the first line in Eq.\@ \eqref{P_e_bayes} as a Gaussian integral and obtain \cite{Guha_PRA_2009}
\begin{equation}
P_e^{(\text{hom})} = \frac{1}{2}\rm{erfc}\left( \sqrt{\frac{\kappa N_s M}{4N_B+2}}\right) \approx \frac{e^{-M\xi_{hom}}}{2\sqrt{\pi M \xi_{hom}}},
\end{equation}
where erfc is the complementary error function, $\xi_{\text{hom}} = \kappa N_s/(4N_B+2)$, and the approximation holds for $\kappa N_s M \gg 4N_B +2$.
In particular, for $N_B \gg 1$, we have $\xi_{\text{hom}}  \approx \kappa N_s/4N_B = \xi_C$. 
Accordingly, homodyne detection saturates the Chernoff bound for a coherent state transmitter.

In quantum illumination, the information on the presence/absence of the target is encoded in the {\it phase-sensitive} cross-correlations (see the covariance matrices \eqref{QIcov}) that depends on both quadratures and therefore cannot be resolved by homodyne or heterodyne detection without incurring in additional noise.
Photon-coincidence counting, which is routinely used to detect the correlations between signal and idler photons as generated in the SPDC process, is also of no use in quantum illumination because of the strong thermal background.
Finally, interferometrically combining the return and the idler modes only allows to access {\it phase-insensitive} correlations. 
As a consequence, it is not possible to exploit the quantum advantage offered by quantum illumination by using standard quantum optics measurements.

To overcome this problem, Guha and Erkmen proposed a receiver that maps phase-sensitive cross-correlation into photon counts \cite{Guha_PRA_2009}.
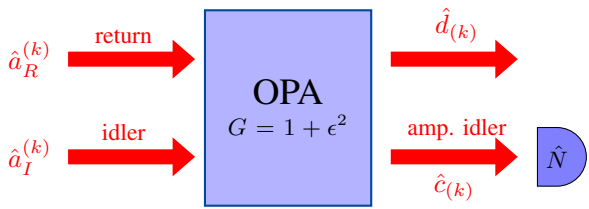
\begin{SCfigure}[][thb]
\centering
\begin{tikzpicture}[scale=1.3]
\tikzset{arw/.style={>={Triangle[length=3mm,width=5mm]},line width=2mm,draw=red}}
\path[draw=MyBlue, thick, fill=blue!30!] (1.4,5.5) -- (1.4,7.5) -- (3.1,7.5) -- (3.1, 5.5) -- cycle; 
\node[text width = 2cm, align=center] at (2.25,6.5) {{\Large OPA}\\{ \small $G=1+\epsilon^2$}};
\draw[arw,->] (0,7) node[left,red] {$\hat{a}^{(k)}_R$}--  node[above,red,xshift=-3] {{\small return}} (1.3,7);
\draw[arw,->] (0,6) node[left,red] {$\hat{a}^{(k)}_I$} --  node[above,red,xshift=-3] {{\small idler}} (1.3,6);
\draw[arw,->] (3.3,6) --  node[above,red,xshift=1] {{\small amp. idler}} node[below,red]{$\hat{c}_{(k)}$}(4.6,6);
\draw[arw,->] (3.3,7) --  node[above,red,xshift=1] {$\hat{d}_{(k)}$} (4.6,7);
\begin{scope}
\clip (5,5.5) rectangle (6,6.5);
\draw [fill=blue!50!] (5, 6) circle(0.3);
\end{scope}
\draw [fill=blue!50!] (5, 5.7) -- (4.8,5.7) -- (4.8,6.3) -- (5,6.3);
\node at (5., 6.) {\small $\hat{N}$};
\end{tikzpicture} 
\caption{Schematic representation of the OPA receiver. The mode returning from the target region and the retained idler are fed to an OPA, with a small gain $G = 1+\epsilon^2$, and the number of photons $N$ in the amplified idler mode is counted. A decision in favour of $H_{0(1)}$ is made if $N$ is smaller (larger) than a threshold value $N_{\rm th}$. }
\label{Fig:OPA}
\end{SCfigure}

This receiver uses an optical parametric amplifier (OPA)\footnote{
We will talk here about OPAs, since this receiver is commonly referred to as an OPA receiver. However, the mathematical description presented here also apply to the microwave case where JPAs are used instead. Note that these are the same kind of non-linear elements used to produced entangled photons.}(see Fig.\@ \ref{Fig:OPA})
that performs the following transformation
\begin{subequations}
\begin{align}
\hat{c}_{(k)} &= \sqrt{G}\hat{a}_I^{(k)} +\sqrt{G-1}\hat{a}_R^{\dagger(k)}\\
\hat{d}_{(k)} &= \sqrt{G}\hat{a}_R^{(k)} +\sqrt{G-1}\hat{a}_I^{\dagger(k)},
\end{align}
\end{subequations}
where $G = 1+\epsilon^2$ is the gain of the OPA, and then counts photons in the amplified idler mode $N_k = \langle\hat{c}^\dagger_{(k)}\hat{c}_{(k)}\rangle$. 
Under both hypotheses the mode $\hat{c}_{(k)}$ is in a thermal state $\rho_c$ given by Eq.\@ \eqref{thermal} with mean photon number \cite{Guha_PRA_2009}
\begin{subequations}
\begin{align}
N_0 &= GN_s + (G-1)(1+N_B),\\
N_1 &= GN_s + (G-1)(1+N_B+\kappa N_s) +2\sqrt{G(G-1)}\sqrt{\kappa N_s(N_s+1)},
\end{align}
\label{N_m}
\end{subequations}
under hypotheses $H_0$ and $H_1$ respectively.

Given $M$ copies of the amplified idler state, the optimal strategy to distinguish between hypotheses $H_0$ and $H_1$ consists in counting the total number of photons $n$ and comparing it with a threshold \cite{Guha_PRA_2009}.
The probability distribution of $n$ under the two hypotheses can be calculated from the quantum state $\rho_c^{\otimes M}$, and yields
\begin{equation}
P_{0/1}(n) =\binom{n+M-1}{n}\frac{(N_{0/1})^n}{(1+N_{0/1})^{n+M}}.
\label{pN_OPA}
\end{equation}
In Eq.\@ \eqref{pN_OPA} the binomial coefficient takes into account the different ways of distributing $n$ photon counts within $M$ measurements, while the fraction gives the probability for one of this way to occur.
This probability distribution has a mean value equal to $MN_{0/1}$ and a variance $M\sigma_{0/1}^2$, with $\sigma_{0/1}^2 = N_{0/1}(N_{0/1}+1)$.
Using the central limit theorem, for $M\gg1$, we can approximate Eq.\@ \eqref{pN_OPA} with a Gaussian function, and then use the Bayesian decision rule \eqref{Bayes_decision_rule} to derive the threshold $N_\text{th} = M(\sigma_1N_0+\sigma_0N_1)/(\sigma_0 +\sigma_1)$, such that we decide that $H_0$ is true if $N < N_\text{th}$ and vice-versa otherwise.
We will come back to this Gaussian approximation later when we will calculate the ROC for the OPA receiver.

We can now evaluate the mean error probability associated with this strategy by calculating the classical Bhattacharyya bound. 
Since $n$ is a discrete random variable, we do this by replacing the integral in Eq.\@ \eqref{classical_Bhattacharyya} with a sum, and we obtain  \cite{Guha_PRA_2009}
\begin{equation}
P_{e,\text{OPA}} \leq \frac{1}{2}(Q_B)^M,
\end{equation}
with 
\begin{equation}
Q_B = \frac{1}{\sqrt{(1+N_1)(1+N_0)} -\sqrt{\vphantom{(}N_0N_1}}
\end{equation}
\begin{SCfigure}
\centering
\includegraphics[width = 0.4\textwidth]{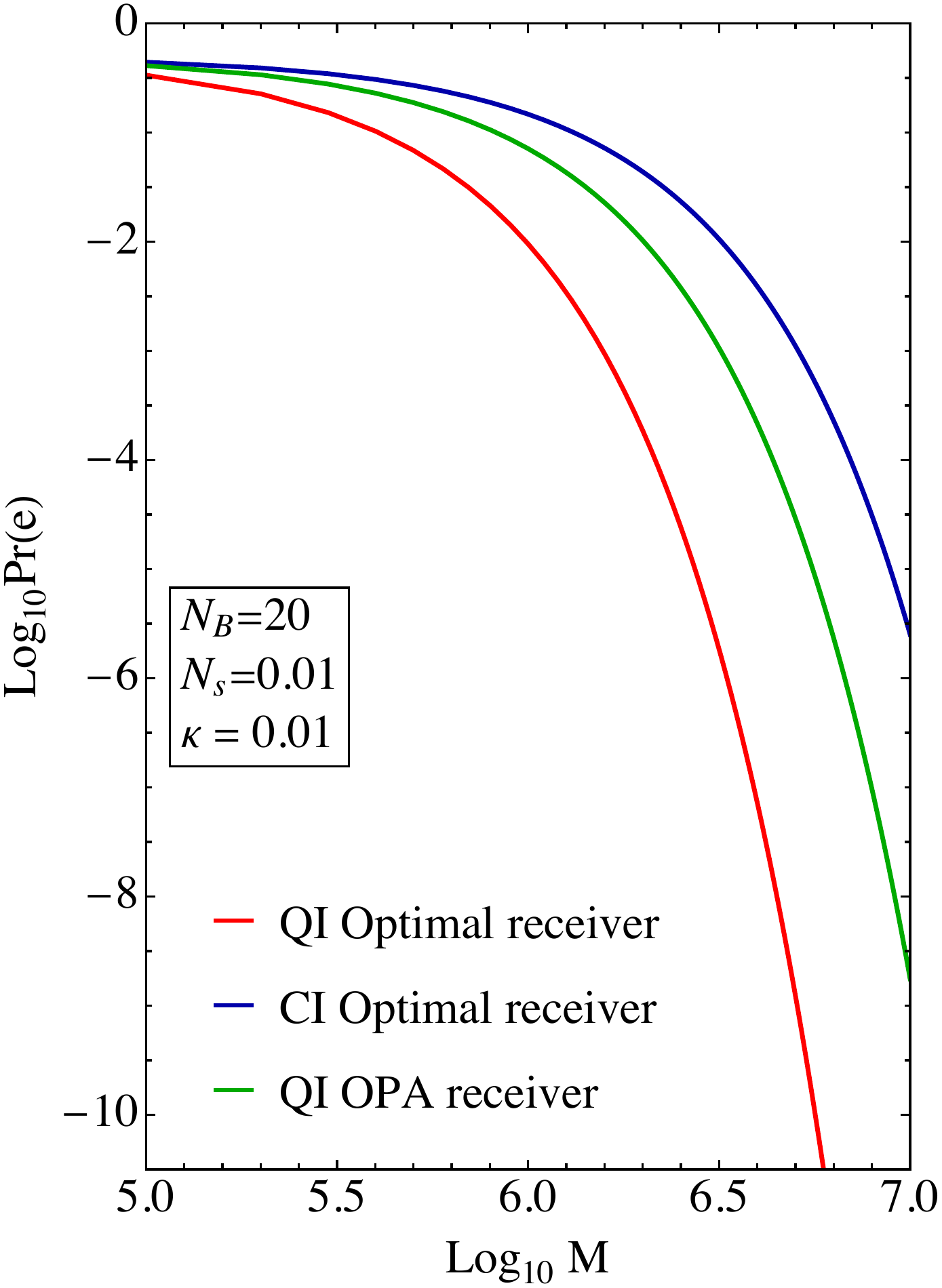}
\caption{Upper bounds for the mean error probability $P_e$ for (blue) classical illumination (Chernoff bound saturated by coherent state transmitter with homodyne receiver), and a quantum illumination transmitter with (green) OPA-receiver (classical Bhattacharyya bound) and (red) ideal receiver (quantum Chernoff bound), as a function of the number of copies $M$ of the state used to interrogate the target region.}
\label{Fig:OPA_plot}
\end{SCfigure}
For a low gain OPA ($\epsilon^2 \ll 1$), we have $N_1 - N_0 = 2\epsilon \sqrt{\kappa N_s(N_s+1)}\ll 1$. We can therefore Taylor-expand $Q_B$ in powers of $N_1 - N_0$, taking into account that we also have $N_1 -N_0 \ll N_0$, and obtain
\begin{equation}
Q_B \approx  1 - \frac{(N_1 -N_0)^2}{8N_0(N_0+1)}  = 1 - \xi_B \approx e^{-\xi_B}.
\end{equation}
Using the expressions for $N_0$ and $N_1$, we have 
\begin{equation}
\xi_B = \frac{\epsilon^2\kappa N_s(N_s+1)}{2N_s(N_s +1) + 2\epsilon^2(2N_s +1)(N_B+N_s+1)},
\end{equation}
which for $N_s \ll 1$, $\kappa \ll 1$, and $N_B \gg 1$ is upper bounded by $\xi_B \approx \kappa N_s/ (2 N_B)$.
This upper bound is reached when $\epsilon^2 N_B \gg N_s$, e.g.\@ when $\epsilon^2 = N_s/\sqrt{N_B}$.
The performances of quantum illumination with the OPA receiver are illustrated in Fig.~\ref{Fig:OPA_plot}. 

The OPA-receiver allows for an error exponent $\xi_B$ which is twice the classical one $\xi_C$, corresponding to $3$ dB quantum advantage, using only off-the-shelf components. 
However, this is $3$ dB less then the ultimate quantum advantage attainable with quantum illumination. 
In the following section, we will discuss how to fill this gap.

\subsection{The ultimate receiver: the feed-forward sum-frequency-generation (FF--SFG) receiver.}
\label{Sec:ultimate}
The sub-optimality of the OPA receiver discussed above stems from the fact that it analyses the $M$ return-idler mode pairs one by one.
This strategy is known to be suboptimal for the discrimination of two mixed quantum states like those described by the covariance matrices \eqref{QIcov} \cite{calsamiglia2010}.
In order to overcome this limit, Zhuang \emph{et al.\@} \cite{Zhuang_PRL_2017} proposed to use SPDC's inverse process: sum frequency generation (SFG).
SFG happens when a pair of photons with frequencies $\omega_S$ and $\omega_I$ and wave vectors $k_S$ and $k_I$ meet on a non-linear device identical to the one used for SPDC, and a photon in the pump mode, with frequency $\omega_P = \omega_S+\omega_I$ and wave vector $k_P = k_S + k_I$ is generated.
Therefore, the receiver Zhuang \emph{et al.\@} proposed in \cite{Zhuang_PRL_2017} consists in going beyond the pair-by-pair analysis of the return and idler modes by combining all $M$ received states that interrogated the target region via SFG such that the information on the presence (absence) of the target gets mapped in the presence (absence) of photons in the pump mode.

\begin{SCfigure}[][h]
\centering
\includegraphics[width = 0.5\textwidth]{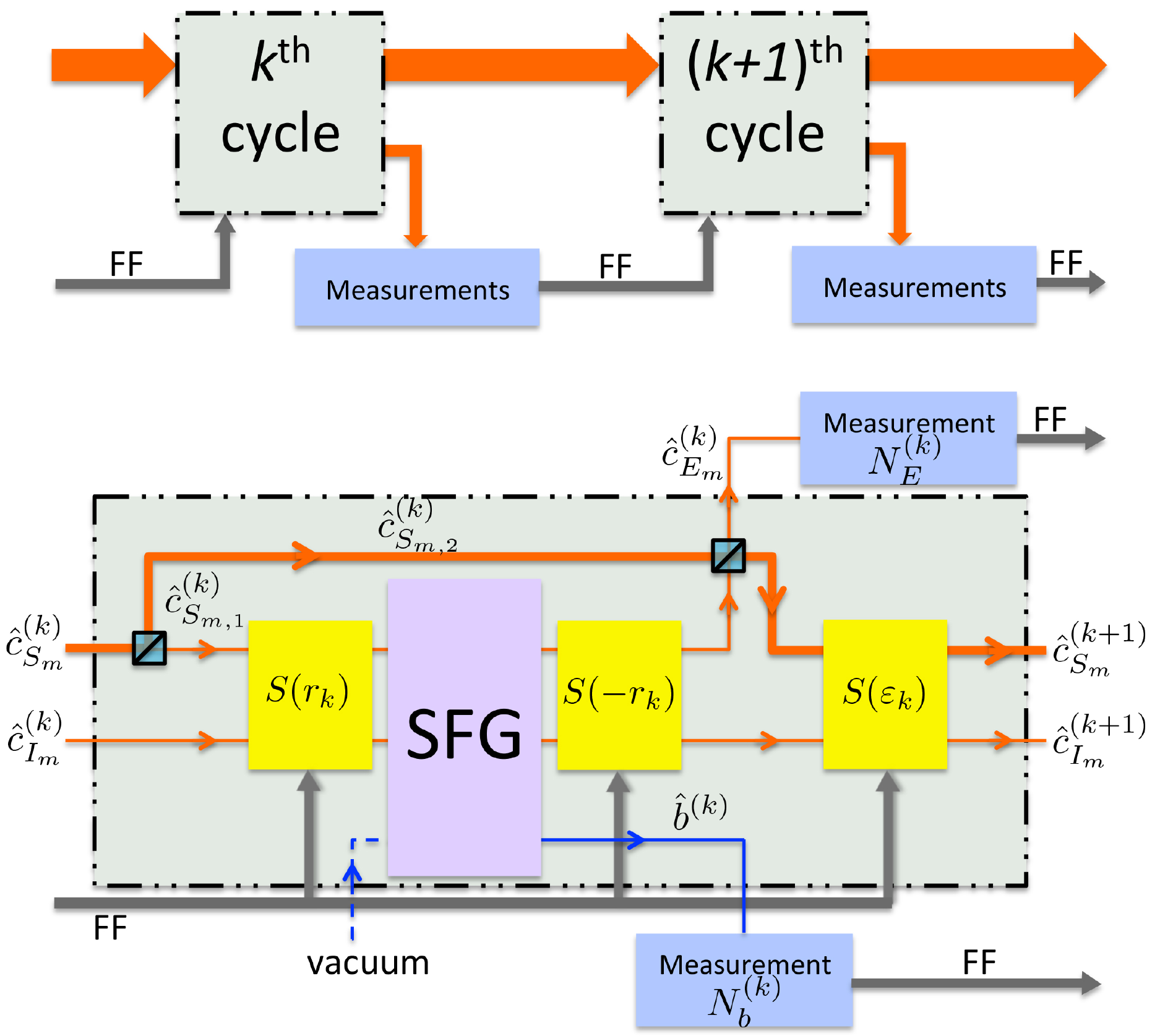}
\caption{(Figure from \cite{Zhuang_PRL_2017}) Schematic  representation of the feed-forward sum-frequency-generation (FF--SFG) receiver. 
The upper part of the figure shows two successive cycles of the feed-forward circuit. 
The lower part shows the structure of the $k$th cycle where a part of return mode is combined with the idler on the SFG device and measurements on the sum-frequency mode and the transformed signal mode are performed.
The yellow blocks represent additional two-mode non-linear operations with tunable parameters which are conditioned on the the results of measurements on the outputs of the previous cycle.}
\label{Fig:SFG_receiver}
\end{SCfigure}

Following this idea, Zhuang \emph{et al.\@} proposed a configuration that uses several cycles of SFG augmented by a feed-forward (FF) circuit that implements additional non-linear operations conditioned on the results of measurements on the output of the previous cycle (see Fig.\@ \ref{Fig:SFG_receiver}).
This combination of SFG with an FF circuit takes the name of FF--SFG receiver.
Under the (currently unrealistic) assumption that SFG has unit efficiency at the single pair level, this receiver has the remarkable property of being able to saturate the Chernoff bound for quantum illumination [see Eq.\@ \eqref{QI_Pe}], and therefore to achieve the full $6$ dB quantum advantage enabled by quantum entanglement.

A detailed description of how this detector works is quite involved and goes beyond the purpose of this review. 
A qualitative description is presented in the caption of Fig.\@ \ref{Fig:SFG_receiver}, and for further details, we refer the interested readers to \cite{Zhuang_PRL_2017, ZhuangJOSAB2017}.
However, a rough approximation of the performances of this device can be obtained in the $N_s \ll 1$ limit, where the output mode of the FF--SFG receiver is found either in the vacuum $\ket{0}$ ($H_0$ is true), or in the coherent state $\ket{\sqrt{N_s \kappa M/N_B}}$ ($H_1$ is true) \cite{ZhuangJOSAB2017}.

\subsection{Performances: ROC for quantum illumination}
\label{sec:performances}
Having introduced the  OPA receiver, which is sub-optimal, but implementable with current technology, and the FF--SFG receiver, which gives the ultimate performance for quantum illumination, we are ready to compare the ROC for quantum and classical illumination \cite{ZhuangJOSAB2017}.

\begin{SCfigure}[][tb]
\centering
\includegraphics[width = 0.5\textwidth]{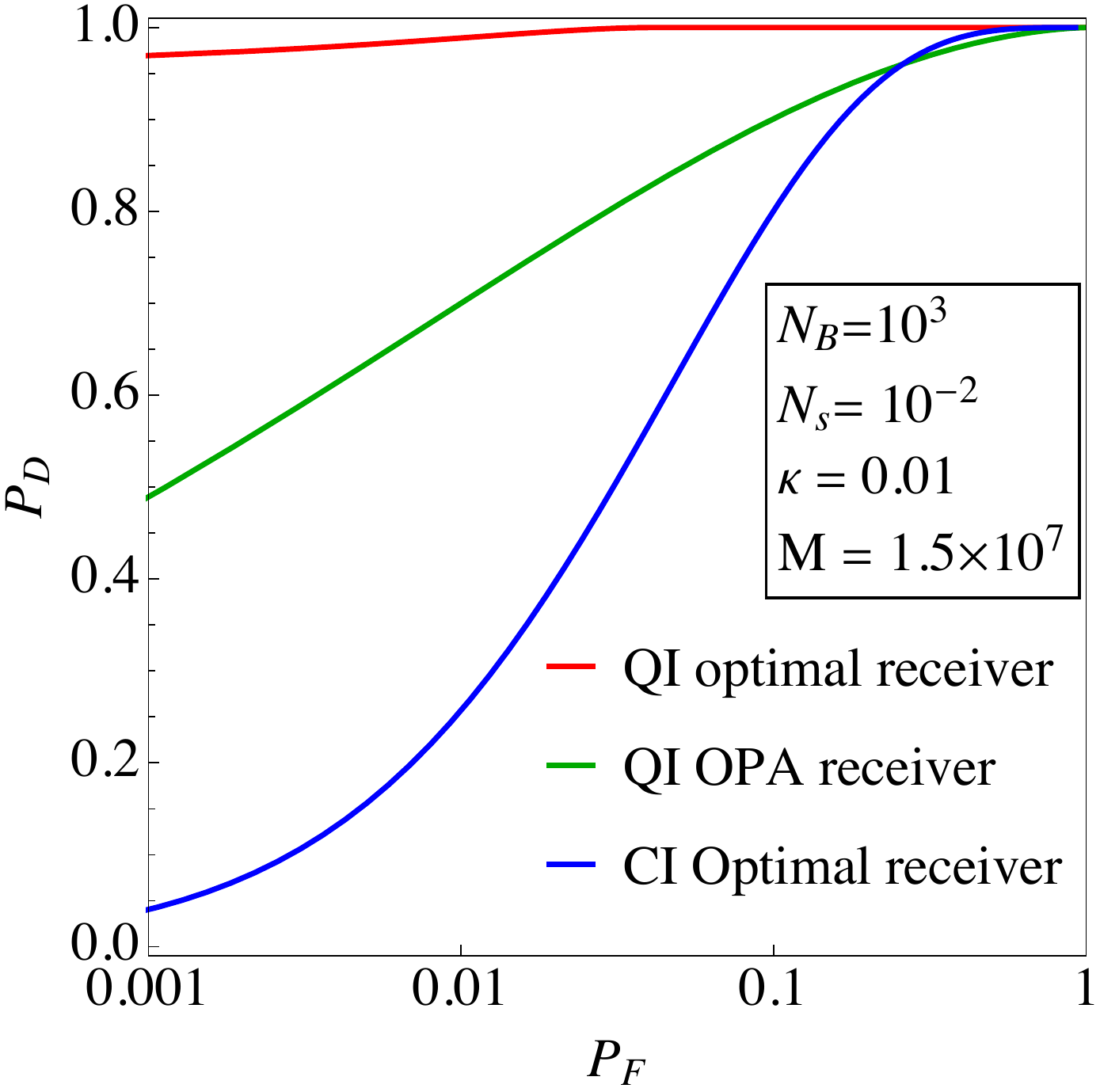}
\caption{ROC: the detection probability $P_D$ as a function of the false alarm probability $P_F$ for different target detection schemes. In blue, the optimal classical illumination (CI) protocol: a coherent-state transmitter with homodyne detection. In green, an entanglement-based scheme realizable with current technology: quantum illumination (QI) with the OPA-receiver. In red, the ultimate limit for quantum illumination, achieved with the FF--SFG receiver. The number of signal photons $N_s$ is chosen in the range where entanglement provides a quantum advantage, while the value of $N_B$ (number of photons in the thermal background) is typical for microwave frequencies.}
\label{Fig:ROC}
\end{SCfigure}

Let us start by considering the coherent state transmitter with homodyne detection. 
In Sec.\@ \ref{sec:receivers}, we have seen that for $N_B\gg1$ this is the optimal classical illumination setting, and that it maps the target detection problem to the discriminations of two Gaussian functions with identical variances, but different means.
In this case, we can derive the probability density for the likelihood ratio $\Lambda$ \cite{VanTrees}, and calculate the false alarm and detection probabilities [see Eqs. \eqref{NP_threshold} - \eqref{NP_PD}] as functions of the threshold $\lambda$ for the Neyman-Person test \eqref{NP_decision_rule}:
\begin{subequations}
\begin{align}
P^{(\text{CI})}_F &= \frac{1}{2} \rm{erfc}\left[\frac{1}{\sqrt{2}}\left(\frac{\log \lambda }{d}+\frac{d}{2}\right)\right],\\
P^{(\text{CI})}_D&= \frac{1}{2} \rm{erfc}\left[\frac{1}{\sqrt{2}}\left(\frac{\log \lambda }{d}-\frac{d}{2}\right)\right],
\end{align}
\label{ROC_coh_hom}
\end{subequations}
with $d = 2\sqrt{M \kappa N_s}/\sqrt{2N_B+1}$. 
From Eqs.\@ \eqref{ROC_coh_hom}, we can plot the ROC for the coherent state transmitter with homodyne detection (blue curve in Fig.\@ \ref{Fig:ROC}).

Let us now move to quantum illumination starting from the OPA receiver (see  Sec.\@ \ref{sec:receivers}).
To calculate the ROC for this receiver, we assume $M\gg 1$, and we use the central limit theorem in order to approximate the photon-counting probability density \eqref{pN_OPA} with a Gaussian function \cite{Guha_PRA_2009}
\begin{equation}
P_{N|H_{0/1}} (n|H_{0/1}) = \frac{e^{-(n-MN_{0/1})^2/(2M\sigma_{0/1}^2)}}{\sqrt{2\pi M\sigma_{0/1}^2}},
\end{equation}
with $N_{0/1}$ defined in Eqs. \eqref{N_m} and $\sigma_{0/1}^2 = N_{0/1}(N_{0/1}+1)$ as discussed in Sec.\@ \ref{sec:receivers}.
We therefore have to calculate the false alarm and detection probability for the discrimination between two Gaussian functions.
In this case, an analytical solution can be obtained using the extended Van Trees receiving characteristic approximation \cite{Shapiro_IEEE_1999}.
The results of this approximation give the green ROC curve in Fig.\@ \ref{Fig:ROC}.

Finally, we consider quantum illumination with its ultimate receiver: the FF--SFG receiver that we discussed in Sec.\@ \ref{Sec:ultimate}.
An approximation of the performance of this receiver is given by the ROC for discriminating the coherent state  $\ket{\sqrt{N_s \kappa M/N_B}}$ from the vacuum.
Given that both a coherent state and the vacuum are pure states, the latter is given by  Eq.\@ \eqref{ROC_pure} with $h = 1-\exp ( - N_s \kappa M/N_B)$ (red curve in Fig.\@ \ref{Fig:ROC}).
This approximation allows us to obtain simple analytical results that describe the performance of what is in fact a complicated, structured receiver. 
The price that we have to pay for this simplification is in the accuracy of such an approximation. 
In particular, as the false alarm probability $P_F \to 0$, the quantum illumination detection probability $P_D$ should also vanish, while the coherent state approximation gives $P_D \to h$  which can be significantly different from zero for modest signal-to-noise ratios $\mathit{SNR} =  N_s \kappa M/N_B$.
A comparison between the approximated ROC for discriminating a coherent state from the vacuum and the one of a numerically simulated FF--SFG is presented in \cite{ZhuangJOSAB2017}.

Fig.\@ \ref{Fig:ROC} shows the performances of quantum illumination, as quantified by the ROC, in the $N_s \ll 1$ regime, where an entanglement-based transmitter provides a significant advantage over a coherent state transmitter.
However, in this regime, since each state contains a very low average photon number per pulse, in order to obtain a detection probability $P_D$ which is close to one, we need millions of copies of the entangled states we use to interrogate the target region (see $M$ in Fig.\@ \ref{Fig:ROC}). 

\begin{SCfigure}[][tb]
\centering
\includegraphics[width = 0.5\textwidth]{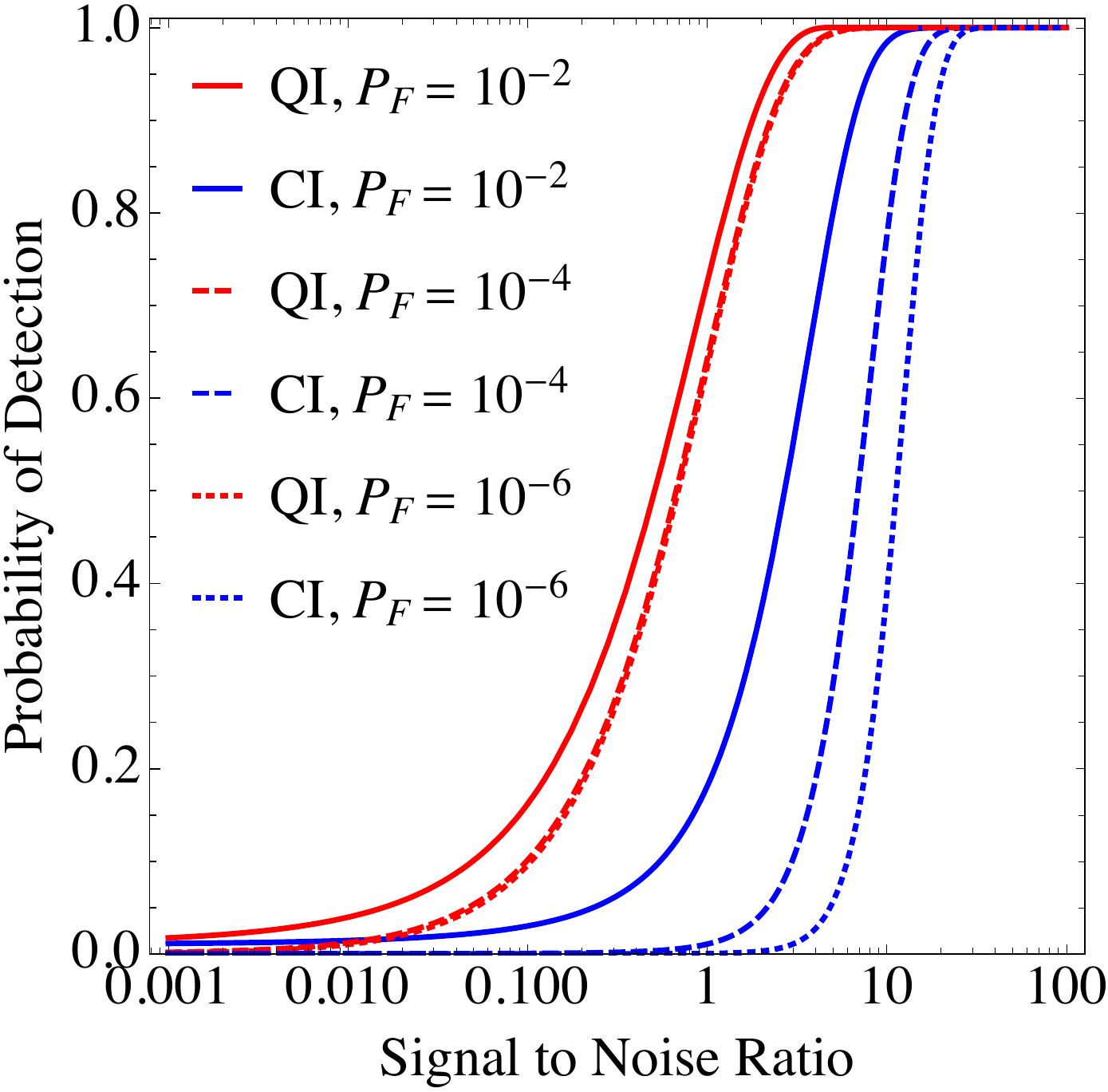}
\caption{Detection probability $P_D$ as a function of the signal-to-noise ratio: $\mathit{SNR} = M \kappa N_s/N_B$, for different values of the false alarm probability $P_F = 10^{-2}$ (solid line), $10^{-4}$ (dashed line) and $10^{-6}$ (dotted line). Blue lines represent the performances of the optimal classical illumination protocols: coherent state transmitter and homodyne detection. Red lines correspond to the quantum illumination with the optimal FF--SFG receiver.}
\label{Fig:SNR}
\end{SCfigure}

To better draw our conclusions, let us now consider a different way to visualize the performance of a target-detection protocol, namely, we plot the probability of detection $P_D$ as a function of the signal to noise ratio ($\mathit{SNR}$) for different fixed values of the probability of false alarm $P_F$\footnote{We thank Roch Settinieri for suggesting us this way to visualize quantum illumination performances.}.
In the present context we can define the $\mathit{SNR}$ as the number of received photons divided by the number of thermal photons: $\mathit{SNR} = M \kappa N_s/N_B$.
It is interesting to notice that the optimal classical illumination performances, as described by Eqs. \eqref{ROC_coh_hom}, depend on the system parameters only through $d$, which for $N_B \gg 1$ can be approximated as $d \approx \sqrt{\mathit{SNR}/2}$. 
Analogously, as discussed in the case of the ROC, the performances of the FF--SFG receiver are well approximated by considering the problem of discriminating the coherent state $\ket{\sqrt{\mathit{SNR}}}$ and the vacuum. 
The performances of these optimal classical and quantum illumination schemes, according to the new metric introduced above, are plotted in Fig.\@ \ref{Fig:SNR}.

Fig.\@ \ref{Fig:SNR} shows an interesting feature of quantum illumination in the Neyman--Pearson scenario, namely the fact that the quantum advantage in this case is not limited to $6$ dB. 
We can see this by noting that the red curves (for quantum illumination) get further apart from the blue ones (for classical illumination) while decreasing the false alarm probability $P_F$. 
Quantitatively, the $SNR$ necessary to achieve an $80 \%$ detection probability with false alarm probabilities $P_F=10^{-2}$, $10^{-4}$ and $10^{-6}$ with quantum illumination (red curves) is smaller than the one in the optimal classical case (blue curves) by, respectively, $6$, $8.2$ and $8.9$ dB\footnote{For some applications an $80 \%$ detection probability may not be enough. If we require a $99 \%$ detection probability, we obtain the more modest quantum advantages of $3.23$, $4.42$, and $4.59$ dB, at $P_F = 10^{-2}, 10^{-4}$, and $10^{-6}$ respectively}.

In fact, by using an entropic approach \cite{Hiai1991,887855}, Wilde \emph{et al.\@} \cite{PhysRevLett.119.120501} proved that in the Neyman--Pearson setting the ratio between the classical and the quantum detection probabilities, i.e the quantum advantage, can be made arbitrary large by making $N_s$ arbitrary small.
However, for $N_s \to 0$, both the above mentioned probabilities go to zero and one needs to consider infinitely many copies of the transmitted state ($M \to \infty$) in order to obtain a finite detection probability.
As we will see in the following, this represents an important drawback of quantum illumination.

\subsection{Criticalities and limitations}
\label{Sec:criticalities}
In this section, we report the main criticalities and limitations of the quantum illumination protocol as they are known in quantum optics literature \cite{pirandola2018, ShapiroReview2019}.
Additionally, we quantitatively characterise the quantum advantage regime in terms of pulse power and duration, in order to give an answer to the question: {\it is quantum illumination practically useful for target detection?}

As discussed in Sec.\@ \ref{sec:gaussian}, the entangled signal-idler pulses needed for quantum illumination are produced in non-linear devices through the process of SPDC, which can happen only for a finite range of frequencies known as the {\it phase matching bandwidth} $W$. 
At optical frequencies ($\omega_s \approx 100$ THz) a typical value of the phase matching bandwidth is $W \approx 1$ THz, while in the microwave regime ($\omega_s \approx 10$ GHz), one usually has $W \approx 100$ MHz.
Accordingly, a pulse of duration $T$ will contain $M = WT$ independent signal-idler modes. 

If we now consider that to have a quantum advantage the number of photons $N_S$ (i.e. the energy $\hbar \omega N_S$) per mode must be very low, in order to reach a significant $SNR = M \kappa N_S /N_B$ the required time-bandwidth product $M=TW$ must be very large. 
In particular, given the bandwidths discussed above, to achieve a time-bandwidth product $M = 10^6$ one would require a pulse duration $T \approx 1\,\mu$s at optical wavelengths, and $T \approx 10$ ms in the microwave range.
Such a long microwave pulse imply that a target will move during the interrogation time inducing a variable time-delay of the return signal. 
This is an issue since to perform the joint quantum measurements described above, the idler must be properly retarded, rotated in polarization, and shifted in frequency in order to match the range, polarization and Doppler bin that one needs to probe.
Accordingly, a target moving over multiple range bins during the interrogation time will cause a temporal mismatch, which induces additional losses that can be quantified by an overlap integral $0 < \kappa_m \leq 1$. 
These losses enter as a multiplicative factor in the signal to noise ratio for a quantum radar, but not for a classical one.
In fact, for a classical (coherent-state) radar a time-bandwidth limited pulse ($M = WT \lesssim 1$) with a high energy per mode ($N_S \gg 1$) achieves the same performances of a pulse with large time-bandwidth product ($M = WT \gg 1$) with a low energy per mode ($N_S \lesssim 1$). Therefore, contrary to its quantum counterpart, classical illumination is not forced to use large time-bandwidth product. 

Another consequence of the high time-bandwidth product requirement of quantum illumination is that the power of these pulses will necessarily be extremely low. 
In fact, the pulse power can be estimated as $P \approx \hbar \omega_s M N_s/T \approx \hbar \omega_s N_s W$, which in the regime where the quantum advantage is observed ($N_S = 0.01$) gives $P \approx 0.01$ fW for microwaves.
The powers of classical radars range from the mWs, used in extremely short-range applications, to the MWs, needed to track planes.
Therefore, the power range where quantum illumination provides a quantum advantage is between $16$ and $20$ orders of magnitude smaller than the typical values used for target detection.
Shorter pulses, and consequently higher powers, could only be enabled by phase-matching bandwidths $W$ far beyond those currently available. 
However, considering that the phase matching bandwidth $W$ cannot be larger than the signal frequency $\omega_s$, the room for improvement is very limited.

The phase-matching bandwidth problem is surely the most severe limitation of quantum illumination.
However, there is another subtlety with the quantum illumination theory presented in Sec.\@ \ref{sec:gaussianQI} that has a strong impact on the practical implementation of the protocol.
In particular, all receivers for quantum illumination described in this tutorial require to store the idler for the whole radar-to-target-to-radar propagation time.
If we consider non-ideal idler storage, all error exponent will be multiplied by a factor $\kappa_I$ equal to the idler transmission coefficient.
On the other hand, in classical illumination, there is nothing to be stored, meaning, for example, that, in the Bayesian setting, $6$ dB of idler storage losses will be enough to destroy the full quantum advantage of quantum illumination.
Storing the idler for time long enough to preserve the quantum advantage poses severe limitation on the range of a quantum radar.

Moreover, in realistic scenarios (especially at optical wavelengths) the amplitude and the phase of the returning light are randomly modified. 
This effect, also known as {\it fading}, nullifies the quantum advantage enabled by the OPA receiver, and  makes the quantum advantage enabled by the FF--SFG receiver sub-exponential \cite{Zhuang_PRA_2017}. 

\section{Experiments on quantum illumination}
\label{Sec:experiments}

We conclude our discussion on quantum radars by presenting what has been achieved so far experimentally at optical and microwave frequencies. A summary of the state of the art of quantum illumination experiments in May 2020 is presented in Table \ref{table}.

The first {\it quantum-illumination-like} experiment was performed by Lopaeva \emph{et al.\@} in 2013 \cite{Lopaeva_PRL_2013}. 
In this experiment, the photon-counting correlations induced by an SPDC source where exploited to obtain an advantage over a correlated-thermal state (correlated-noise radar).
We used the adjective {\it quantum-illumination-like} to describe this experiment because it didn't exploit entanglement.
As a consequence, the Lopaeva \emph{et al.\@} setup could only outperform a correlated-thermal state of the same energy, and not a coherent state transmitter.

In 2015, Zhang \emph{et al.\@} \cite{Zhang_PRL_2015} performed the first real quantum illumination experiment using the protocol from Tan \emph{et al.\@} \cite{Tan_PRL_2008} described in Sec.\@ \ref{sec:gaussianQI} together with the OPA receiver \cite{Guha_PRA_2009} discussed in Sec.\@ \ref{sec:receivers}.
Due to experimental imperfections, this experiment couldn't achieve the full $3$ dB quantum advantage, but it demonstrated a 20\% (0.8 dB) enhancement of the error probability exponent.
Up to today the  work of Zhang \emph{et al.\@} remains the only experiment where a quantum illumination protocol has been compared with an optimal classical setup of the same power, and a quantum advantage has been demonstrated.

A more recent experiment, by England \emph{et al.\@} \cite{EnglandPRA2019}, used entanglement and photon-coincidence counting. 
However, the authors operated in a low background regime, where no quantum advantage is possible.

While all three experiments mentioned above were performed at optical frequencies, the recent development of new sources of microwave entanglement enabled the first quantum radar experiments in this frequency domain \cite{luong2019quantum,luong2019receiver,Chang_applied_phys_lett_2019,Barzanjeh_arxiv_2019}.
All these works used JPA to produce entangled photons in the GHz regime, and amplified the signal and idler before sending it to the target region.
However, none of these experiments implement the quantum illumination scheme discussed in Sec.\@ \ref{sec:gaussianQI}.
In fact, instead of performing joint measurements on the stored idler and the signal coming back from the target region, they heterodyne-detect the idler immediately after amplification, and then compare it digitally with the heterodyne-detected return mode.

All these works present comparisons with some classical radars, and show that their quantum devices outperform them.
However, all the considered classical and quantum radars are not optimal, and therefore these works cannot be considered a proof of quantum advantage.
In fact, there are two important flaws in the above described procedure that prevent all these experiment to demonstrate a true quantum advantage.

The first of these flaws is that by heterodyne-measuring the return and idler modes one introduces some additional noise that deteriorates the correlations between the two modes \cite{Weedbrook_Rev_Mod_Phys_2012}.
There are very efficient classical strategies to counteract this noise which however cannot be applied to the quantum case \cite{ShapiroReview2019}. 
As a result, when the signal and return mode are measured individually via heterodyne detection it is always possible to find a classical radar that performs as well (sometimes even better) than the quantum one.
Furthermore, even an ideal heterodyne measurement of the idler would only project the signal beam onto a coherent state \cite{Weedbrook_Rev_Mod_Phys_2012, 1056132}, implementing de facto a (suboptimal) classical illumination strategy as defined in Sec.\@ \ref{sec:gaussianQI}.
The only way to exploit the benefits of quantum correlations is to perform a joint measure on the return and idler modes as in the cases of the OPA and FF--SFG receivers \cite{Weedbrook_NJP_2016}.
The only microwave quantum experiment that discusses this issue is the one by Barzanjeh \emph{et al.\@} \cite{Barzanjeh_arxiv_2019}, where the experimental data are used to simulate a joint-measurement scenario and to prove that in that case it is possible to show a quantum advantage.

The second criticality of these microwave experiments is represented by the pre-amplification of signal and idler before interrogating the target region.
In fact, quantum mechanics ensures that amplification with gain $G$ always come with some noise of variance $G-1$, that reduces the entanglement of the signal-idler pair.
In the experiment by Barzanjeh \emph{et al.\@} \cite{Barzanjeh_arxiv_2019} the pre-amplification noise was actually large enough to disentangle the signal and the idler before interrogating the target region. 
In this case, as admitted by the authors themselves, the strongest signature of the target presence is given by the amplifier noise, and not by the correlations between the signal and idler. 
Accordingly, quantum illumination is not practically relevant in this scenario.

{ 
\nohyphens{\sloppy
\renewcommand{\arraystretch}{1.5}
\begin{table}[t]
\centering
\begin{tabularx}{\textwidth}{>{\Centering\arraybackslash}m{3.cm}*{5}{>{\Centering\arraybackslash}X}}
\toprule
{\bf Experiment} &{\bf Lopaeva \emph{et al.\@}} \cite{Lopaeva_PRL_2013} & {\bf Zhang \emph{et~al.\@}} \cite{Zhang_PRL_2015} & {\bf Luong, Chang \emph{et al.\@}} \cite{luong2019quantum,luong2019receiver,Chang_applied_phys_lett_2019}  & {\bf England \emph{et al.\@}} \cite{EnglandPRA2019} & {\bf Barzanjeh \emph{et al.\@}} \cite{Barzanjeh_arxiv_2019}\\
\midrule
 Frequency  & optical & optical & microwave  & optical & microwave \\
$N_s \ll 1 $& \textcolor{MyGreen}{\cmark} &\textcolor{MyGreen}{\cmark} & \textcolor{MyGreen}{\cmark}& not specified & \textcolor{MyGreen}{\cmark}\\
$ N_B \gg 1 $ & \textcolor{MyGreen}{\cmark} & \textcolor{MyGreen}{\cmark}& \textcolor{MyGreen}{\cmark} &\textcolor{MyRed}{\xmark}&\textcolor{MyGreen}{\cmark}\\
optimal classical setup & \textcolor{MyRed}{\xmark} & \textcolor{MyGreen}{\cmark} & \textcolor{MyRed}{\xmark} & \textcolor{MyRed}{\xmark} & \textcolor{MyRed}{\xmark}\\
joint measurement& \textcolor{MyRed}{\xmark} & \textcolor{MyGreen}{\cmark} & \textcolor{MyRed}{\xmark} & \textcolor{MyGreen}{\cmark} & \textcolor{MyRed}{\xmark}\\
quantum advantage& \textcolor{MyRed}{\xmark} & \textcolor{MyGreen}{\cmark} 20\% (0.8 dB)& \textcolor{MyRed}{\xmark} & \textcolor{MyRed}{\xmark}& \textcolor{MyRed}{\xmark}\\
\toprule
\end{tabularx}
\caption{State of the art of quantum illumination experiments.}
\label{table}
\end{table}
}
}

\section{Conclusion}
In this review, we explained to an audience not necessarily familiar with quantum optics how quantum entanglement can be used to improve target-detection performances in presence of high losses and a strong thermal background.
In particular, we showed that in the low signal limit,  quantum illumination allows for a reduction of the error probability. 
At the same time, the quantum advantage becomes less and less important when the number of photons per mode used to interrogate the target region is increased.

If we look closer at the physical processes that can be used to produce entangled photons, in particular in the microwave regime, the pulse powers available at the source in the region of quantum advantage are extremely low: 15 to 20 orders of magnitude below what is typically used in radar applications.
As a consequence, in practical scenarios where the radar-to-target round-trip transmissivity is incredibly low (up to $10^{-20}$ for plane tracking applications) in order to have a signal to noise ratio corresponding to a non-negligible detection probability one would need very unpractically long pulses.  
One could hope to fill this huge gap with an even larger quantum advantage, obtained for example by adding additional entanglement between the $M$ temporal modes composing the signal and idler pulses. However, recent works from De Palma and Borregard \cite{DePalma_2018} and Nair and Gu \cite{Nair:20} proved that the $M-$modes two-mode squeezed vacuum used in the Gaussian quantum illumination protocol from Tan {\it et al.} \cite{Tan_PRL_2008} is essentially quantum optimal for target detection.
Therefore, while it is clear that one decade of research in this field provided us with several academically interesting results, their practical relevance seems to be limited.

Nevertheless, quantum illumination represents the first example of an entanglement-based protocol that provide a quantum advantage even though the initial entanglement is completely destroyed by noise and losses in the system.
In other words, we should not dismiss the value of quantum entanglement even in those adverse environmental conditions (severe losses and high noise level) which are often present at microwave frequencies.
Moreover, concepts from quantum illumination can be relevant for other fields, for example, the OPA \cite{Guha_PRA_2009} and the FF--SFG \cite{Zhuang_PRL_2017} receivers recently found applications in entanglement-assisted communication schemes \cite{shi2020practical, guha2020infinite}, and it is not to exclude, that they may lead to the discovery of new quantum protocols of real practical utility.

\section*{Acknowledgement}
We would like to thank Quntao Zhuang, Stefano Pirandola, Mark Wilde, and Roch Settinieri for their critical comments which significantly contributed to improve our review. We also thank Saikat Guha and Benjamin Huard for our enjoyable and informative discussions.

\bibliography{Quantum_radar_bib}{}
\bibliographystyle{ieeetr}
\end{document}